\documentclass[final]{elsart} 

\usepackage{graphicx}
\graphicspath{{figures/}}
\usepackage{amsmath}
\usepackage{amssymb}
\usepackage{amsfonts}
\usepackage{url}
\usepackage{fp}

\newcommand{\ip}{IP}
\newcommand{\h}{r}
\newcommand{\traceroute}{trace\-route}
\newcommand{\onelab}{Paris \traceroute}

\newcommand{\opphroute}{measured route}

\newcommand{\nloops}{$n$-loops}

\long\def\symbolfootnote[#1]#2{\begingroup%
\def\thefootnote{\fnsymbol{footnote}}\footnote[#1]{#2}\endgroup}

\renewenvironment{thebibliography}[1]{%
  \begin{oldthebibliography}{#1}%
    \setlength{\parskip}{0ex}%
    \setlength{\itemsep}{0ex}%
  }%
  {%
  \end{oldthebibliography}%
}

\begin{document}

\begin{frontmatter}

\title{Detection, Understanding, and Prevention\\ of Traceroute Measurement Artifacts}
\author[lipsix]{Fabien Viger} 
\author[lipsix]{Brice Augustin}
\author[lipsix]{Xavier Cuvellier}
\author[lipsix]{Cl\'{e}mence Magnien}
\author[lipsix]{Matthieu Latapy\corauthref{ml}}
\ead{Matthieu.Latapy@lip6.fr}
\author[lipsix]{Timur Friedman}
\author[lipsix]{Renata Teixeira}
\address[lipsix]{Universit\'{e} Pierre et Marie Curie -- CNRS, Laboratoire LIP6}
\corauth[ml]{Corresponding author. Address: LIP6 -- 104 avenue du Pr\'esident Kennedy --  75016 Paris -- France\\
Tel: +33 (0)1 44 27 87 84, Fax: +33 (0)1 44 27 74 95}
\date{} 

\begin{abstract}
  Traceroute is widely used, from the diagnosis of network problems to
  the assemblage of internet maps.  Unfortunately, there are a number
  of problems with traceroute methodology, which lead to the inference
  of erroneous routes.
This paper studies particular structures arising in nearly all traceroute
measurements. We characterize them as ``loops'', ``cycles'', and ``diamonds''.
We identify load balancing as a possible cause for the appearance of
{\em false} loops, cycles, and diamonds, 
i.e., artifacts that do not represent the internet topology.
We provide a new publicly-available traceroute,
  called \textit{\onelab{}}, which, by  controlling the packet header
  contents, provides a truer picture of the  actual routes that
  packets follow.
We performed measurements, from the perspective of a single source
tracing towards multiple destinations,
and \onelab{} allowed us to show that many of the particular structures
we observe are indeed traceroute measurement artifacts.
\end{abstract}



\end{frontmatter}

\section{Introduction}\label{Introduction}

Jacobson's \textit{traceroute}~\cite{jacobson1989traceroute} is one
of the most widely used network measurement tools. It reports an IP
address for each network-layer device along the path from a source to
a destination host in an IP network. Network operators and researchers
rely on traceroute to diagnose network problems and to infer
properties of IP networks, such as the topology of the internet.
This has led to an impressive amount of work in recent years
\cite{govindan2000heuristics,huffaker2002topology,spring2002measuring,Cheswick00,dimes,faloutsos99powerlaw,magoni2005mapping},
in which traceroute measurements play a central role.

Some authors have noticed that traceroute suffers from deficiencies that lead
to the inference of inaccurate routes,
in particular in the presence of load balancing
routers~\cite{huffaker2002topology,spring2002measuring,moors2004streamlining}.
However, no systematic study of these deficiencies has been undertaken.
Therefore, people dealing with traceroute measurements currently have no choice
but to interpret surprising features in traceroute measurements
as either characteristics of the routing or of the network's topology.
This is supported by the common assumptions that these
deficiencies have a very limited impact,
and that, in any case, nothing can be done to avoid them.

The core contribution of this paper,
which is a longer version of our earlier work~\cite{augustin2006paris},
 is to show that both of these assumptions
are false.
We show that the wide presence of load-balancing routers
in the internet induces
a variety of artifacts in traceroute measurements, 
and we provide a rigorous approach to both quantify and avoid many of them.

More precisely, we focus on three particular structures often
encountered in traceroute measurements, which
we categorize as ``loops'', ``cycles'', and ``diamonds''.
Using measurements from a single source tracing towards multiple
destinations, we show that many instances of these structures are actually
measurement artifacts resulting from load-balancing routers.
We provide a new traceroute,
called \textit{\onelab},\,\footnote{\onelab{} is free, open-source software, available from \url{http://www.paris-traceroute.net/}.} which controls packet
header contents to largely limit the effects of load balancing,
and thus obtain a more precise picture of the actual routes.
We show that many of the observed structures disappear when one uses \onelab.
Finally, we explain most
other instances using additional information provided by \onelab,
and suggest possible causes for the remaining ones.

Throughout this paper, we use data obtained by tracing routes from one particular
source to illustrate our results (see Sec.~\ref{measurement_setup}).
From the outset, we insist on the fact that this data is not meant to be 
statistically representative
of what can be observed on the internet in general:
it serves as an illustration only,
and the quantities reported may differ significantly from what
would be observed from other sources.
Obtaining a representative view of the average behavior of the
traceroute tool would clearly be of interest,
but is out of the scope of this paper:
we focus here on the identification of traceroute artifacts,
their rigorous interpretation, and their suppression using \onelab{}.

This paper is structured as follows. Sec.~\ref{better}
describes the classic traceroute tool, its deficiencies,
and the new tool we built to circumvent these deficiencies, \onelab{}.
Sec.~\ref{measurement_setup} describes our methodological framework.
Sec.~\ref{anomalies}
categorizes the particular structures that we encounter and study in 
traceroute measurements. Sec.~\ref{controlled_tuples} and 
Sec.~\ref{other_causes} study these structures, and provide our 
explanations for them. Sec.~\ref{related_work} discusses related work.
Finally, Sec.~\ref{conclusion} presents our conclusions and perspectives 
for future work.

\section{Building a better traceroute}\label{Methodology}
\label{better}

This section describes the tools  used in this paper to study traceroute
measurement artifacts.
Sec.~\ref{traceroute} describes the classic traceroute tool,
and may be skipped by those familiar with it.
Sec.~\ref{tr_lb} describes the deficiencies in classic traceroute
in the face of load balancing.
Sec.~\ref{onelab} then presents our new traceroute, \onelab{},
which avoids some of these deficiencies,
notably the ones induced by per-flow load balancing.

\subsection{Traceroute}\label{Traceroute}\label{traceroute}

There are many varieties and derivatives of the traceroute tool.  
This section describes a generic probing scheme,
based on Jacobson's version~\cite{jacobson1989traceroute}.
Related tools,
such as \textit{NANOG traceroute}~\cite{gavron1995nanog}
and \textit{skitter}~\cite{huffaker2002topology}, do very similar things. 
For a good detailed description of how traceroute works, see
Stevens~\cite{stevens1994traceroute}.

At a high level, three parameters define an invocation of the
traceroute tool: the destination IP
address, $d$, the probe packet protocol, $P$, and the number of probes
per hop, $n$. 
The address $d$ may be any legal IP address.  The protocol
$P$ is one of either: UDP (by default), ICMP (also fairly common), or
TCP (not used by the classic traceroute, but more and more used by other tools,
such as \textit{tcptraceroute}~\cite{toren2001tcptraceroute},
because there is often less filtering of TCP packets).

Probing with traceroute is done hop by hop, moving away from the source
towards the destination in a series of rounds.
Each round is associated with a {\em hop count}, $h$.
The hop count starts at one and is incremented after each round
until the destination is reached, or until another stopping condition
applies.
A \textit{round} of probing consists in sending $n$ probe
packets with protocol $P$ towards destination $d$. By default $n$ is equal to three.
The probe packets are sent with the value $h$ in the IP time-to-live (TTL) field.

The TTL of an IP packet is supposed to be examined by each router that
the packet reaches.  If the TTL is greater than one, then it is decremented and
the packet is forwarded.
If it  is equal to one,
the router  drops the packet and 
sends an ICMP \textit{Time Exceeded} message back to the
source~\cite{ip-rfc791}. 
Routers are required to employ, as the source address of this ICMP message,
the address of the IP interface that sends the ICMP
packet~\cite[Sec.~4.3.2.4]{routers-rfc1812}.\,\footnote{For more
details, see Mao et al.~\cite{mao2003astr} and references within.}
When routing is symmetric, this is typically the address of the interface 
that received the probe.
The reception of a {\em Time Exceeded} message allows the traceroute 
tool to infer the presence of the source address at distance $h$ on the path 
to $d$.

The traceroute tool requires some means of matching
return packets with the corresponding probe packets to know the
correct sequence of IP addresses in the route to $d$.
This is done by examining the payload of the {\it Time Exceeded} packet,
which contains the beginning of the probe packet.
More precisely, an ICMP \textit{Time Exceeded} message contains the IP header of the probe packet,
and the first eight octets that follow the IP header~\cite[p.5]{rfc792}.
If the IP header contains no options, as is the
case for traceroute probe packets, this amounts to the first 28 octets of the IP packet.
The eight octets following the IP header comprise
either the entirety of the UDP header, the entirety of the ICMP \textit{Echo} header 
(the standard four octets of the ICMP header plus four octets for the Identifier and Sequence
Number fields), or part of the TCP header,  depending on $P$.
If this portion of the probe packet contains a unique identifier,
then traceroute can recognize the identifier
in the {\it Time Exceeded} packet and match it to the corresponding probe.
The default behavior of traceroute consists in
setting the Source Port value of UDP probes
to the running process identifier (PID) plus 32,768, and the initial
Destination Port value to 33,435, and incrementing it with each probe sent.
For ICMP {\em Echo} probes, the Sequence Number field is incremented
with each probe sent~\cite{jacobson1989traceroute}.

For various reasons (such as routers dropping probes with a TTL of 1 without
notifying the source, or routers on the reverse path dropping ICMP {\it Time
Exceeded} messages), there might be no answer to a given probe.  If, 
after a given time interval has elapsed, traceroute has not received an 
answer for a given probe, it stops waiting for it and outputs a star (`*') 
for the corresponding probe.

\subsection{Traceroute and load balancing}
\label{load balancing}
\label{tr_lb}

Network administrators employ load balancing to enhance reliability
and increase resource utilization. The main way to do so is through the intra-domain
routing protocols OSPF~\cite{ospf-rfc2328} and IS-IS~\cite{isis-rfc1195} 
that support~\textit{equal cost multipath} (\textit{ECMP}). An operator of a multi-homed 
stub network can also use load balancing to select which of its 
internet service providers will receive which packets~\cite{quoitin03interdomain}.

Routers can spread their traffic across multiple equal-cost paths
using a per-packet, per-flow, or per-destination 
policy~\cite{cisco-lb,juniper-lb}.  In \textit{per-flow load balancing},
packet header information ascribes each packet to a flow, and the
router forwards all packets belonging to a given flow to the same
path.
This helps to avoid packet reordering within a flow.
\textit{Per-packet load balancing} makes no
attempt to keep packets from the same flow together, and focuses
purely on maintaining an even load on paths.
  This might be through round-robin
assignment of packets to paths.  
The balance cannot be disturbed by the presence of
out-sized flows.
Finally, \textit{per-destination load balancing} could be seen as a
coarse form of per-flow load balancing,
as packets are directed as a function of their destination IP address.
But, as it disregards source information,
there is no notion of a flow {\em per se}.  As seen from the measurement point of
view, per-destination load balancing is equivalent to classic
routing, which is also per destination.

Concerning per-flow load balancing, a natural flow identifier is the
\textit{five-tuple} of fields from the IP header and either the TCP or
UDP header: Source Address, Destination Address, Protocol, Source
Port, and Destination Port.
We performed some tests on some load-balancing routers from our traces to get an
indication of which fields are used by routers to determine whether
two packets belong to a same flow or not.  We used TCP, UDP, ICMP,
and IPSec probes.
We sent probes from our laboratory to different destinations that cross the selected routers and varied
header fields to observe which ones triggered load balancing.
These tests showed that routers balance load using 
various combinations of the fields of the five-tuple, as well as three 
other fields: the IP Type of Service (TOS), and the ICMP Code and 
Checksum fields.  
We have not obtained answers from routers for ICMP probes of
any type other than ICMP {\em Echo}, which means that we
were unable to ascertain whether the ICMP Type field 
is also used for per-flow load balancing or not.
Fig.~\ref{fig:headers} summarizes the IP,
UDP, ICMP {\em Echo}, and TCP header fields that we observed are used
by per-flow load balancers.  We leave an
exhaustive study of which parts of the header  fields serve for load balancing, and
in precisely which ways, to future work.

Finally, whether a router balances load per-packet, per-flow or
per-destination depends on the router manufacturer, the OS version,
and how the network operator configures it.
For instance, Cisco and Juniper routers can be configured to do any
of the three types of load
balancing~\cite{cisco-lb,cisco-pf,juniper-lb}.

Traceroute, in consequence of its design, systematically sends probes
via different paths in the presence of per-flow load balancing.
This comes from its manipulation of the contents of the 28 first
octets of the probes, in order to obtain a unique identifier.
When sending UDP probes, it systematically
varies the Destination Port field.
When sending ICMP {\em Echo} probes, it varies the Sequence Number field.
However, as explained above, varying these fields
amounts to changing the flow identifier for each probe.
The Destination Port field
in the UDP header is used for flow identification,
and, though the Sequence Number field is not directly used in this way,
varying this field
varies the Checksum field, which is a flow identifier.

\medskip
Where there is load balancing, there is no longer a single route from
a source to a destination.
Classic traceroute is not only unable to uncover all
routes from a source to a given destination,
but it also proves  unable to identify one single route from among many.
It suffers from two systematic problems:
it fails to discover routers and links, and it may uncover false links.

\begin{figure}
\begin{center}
\includegraphics[scale=0.4]{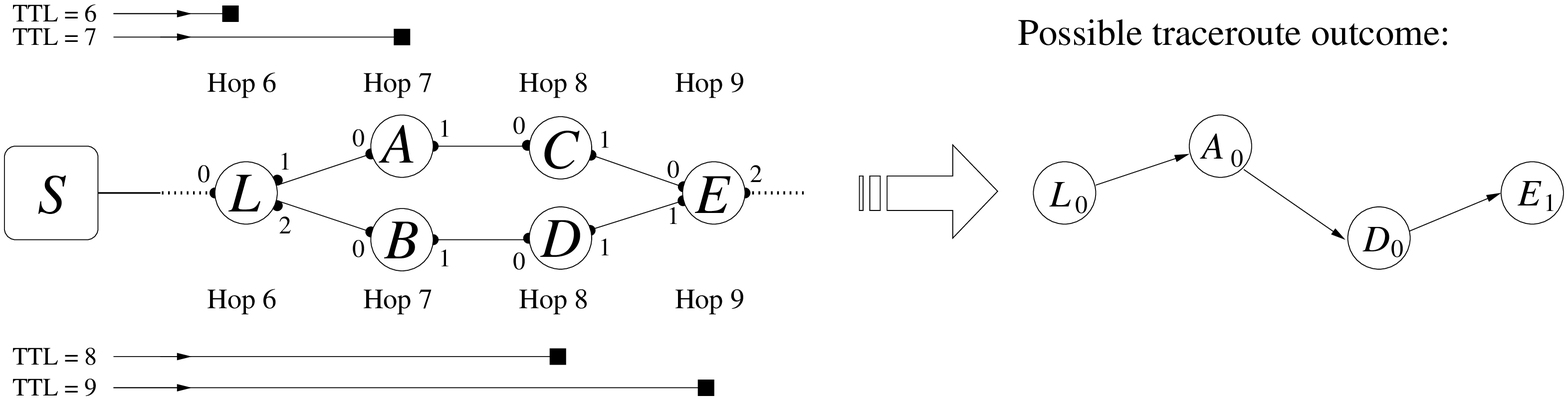}
\end{center}
\caption{Missing routers and links, and false links.}
\label{fig:false-link}
\end{figure}

This is illustrated in
Fig.~\ref{fig:false-link}. Here, $L$ is a load balancer
at hop 6 from the traceroute source, $S$.  On the left,
we see the true router
topology from hop 6 to hop 9. Circles represent routers,
and each router interface is numbered. 
Black squares depict probe packets sent with TTLs 6 through 9.
They are shown either above the 
topology, if $L$ directs them to router $A$, or below, if $L$ directs them 
to router $B$. On the right, we see the topology that would
be inferred from the routers' responses. 

\textbf{Missing routers and links.}  Because routers $B$ and $C$ send
no response, they are not discovered, and links such as $(L_0,B_0)$
and $(B_0,D_0)$ cannot be inferred. The
probability of missing routers goes up with the number of routers at a
given hop count: for instance, if there are four routers at a given hop,
assuming that all probes have an equal  probability of reporting each of the four routers,
the probability of missing at least one of them
when sending four probes 
is approximately equal to 0.78.  Notice also that where there are
missing routers, there are necessarily missing links as well.
Finally, notice that the number of routers at a given hop count may be
quite high: the newer Juniper routers, for instance, permit up to
sixteen equal-cost paths.

\textbf{False links.} 
Independently of whether all routers (or links) are discovered
or not,
the traceroute tool lends itself to the inference of false links.
In our example (Fig.~\ref{fig:false-link}),
$L$ directs the probe with initial TTL 7 to $A$ and the one with initial TTL 8 to $B$,
leading to the mistaken inference of a link between $A_0$ and $D_0$.

In our example, there is a probability $2/2^4 = 0.125$ that 
all probes to hops $7$ and $8$ follow an identical path.
Thus, there is a probability of $0.875$ that addresses from routers on
different paths will be revealed on subsequent hops,
leading to false links impossible to differentiate from true ones in such measurements.
Again, the presence of a high number of routers at a given hop
count complicates the problem further by increasing the probability
of inferring false links.

The problem of false links inferred from the appearance
of multiple addresses
on a same hop has been acknowledged in some cases,
in particular by
Huffaker et al.~\cite{huffaker2002topology} and
Spring et al.~\cite{spring2002measuring}.
However, no systematic study of this phenomenon has been undertaken,
and no real solution has been proposed.

\subsection{A new traceroute}\label{a new traceroute}
\label{onelab}

We now introduce {\em \onelab{}},\,\footnote{Available at \url{http://www.paris-traceroute.net/}}
a new traceroute designed for networks with load-balancing routers. 
Its key innovation is to control the probe packet header fields in a manner
that makes all probes towards a destination follow the same path in
the presence of per-flow load balancing.
It also makes it possible to distinguish between the
presence of per-flow load balancing and per-packet load balancing,
as we see below.
Unfortunately, due to the random nature of per-packet load balancing,
\onelab{} cannot perfectly enumerate all paths in
all situations.
But it can do considerably better than the classic
traceroute, and can flag those instances where there are
doubts.

\begin{figure}[!ht]
\begin{center}
\includegraphics[scale=0.45]{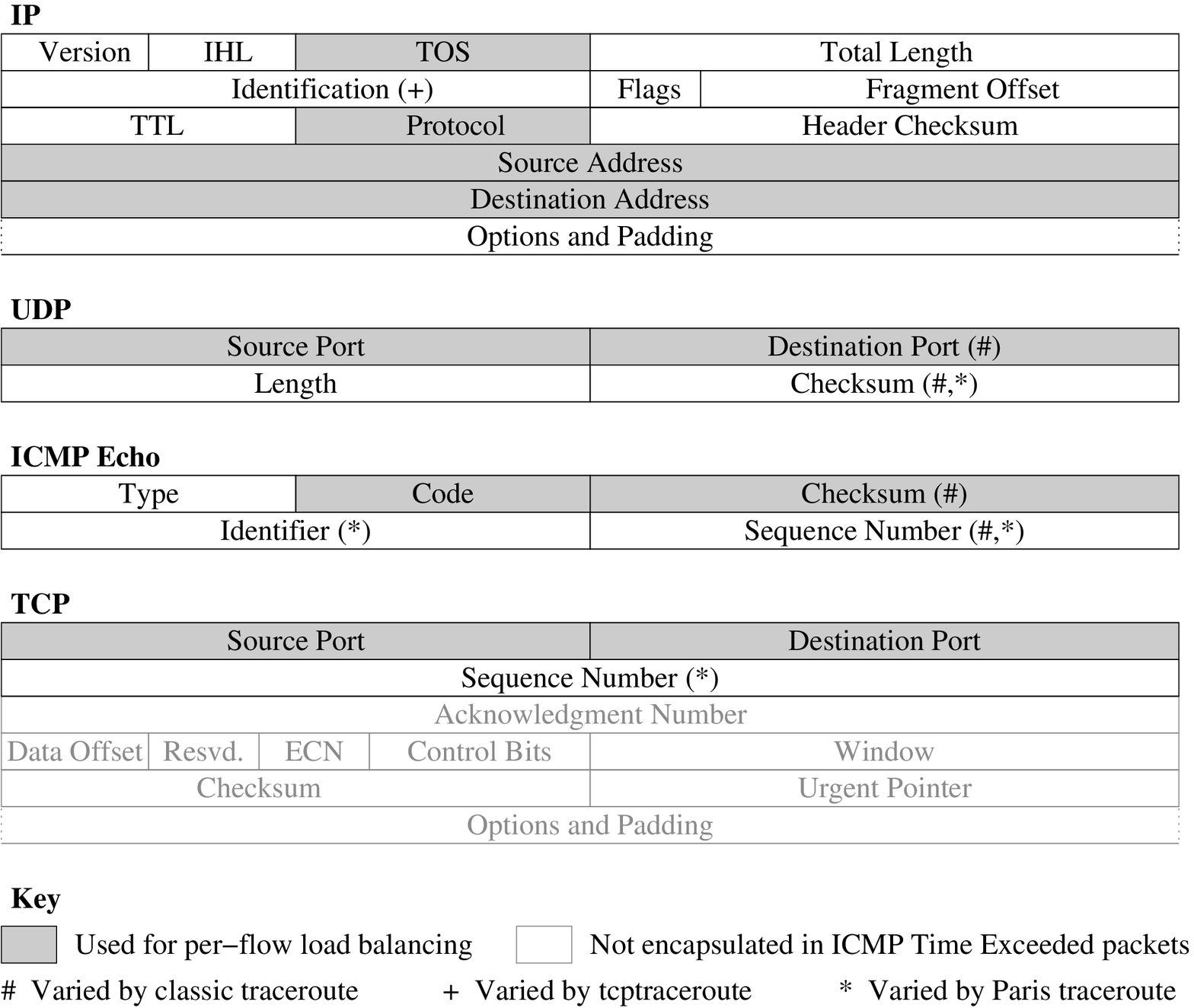}
\end{center}
\caption{The roles played by packet header fields.}
\label{fig:headers}
\end{figure}

Maintaining certain header fields constant is difficult because
traceroute needs to match response packets to their
corresponding probe packets, and,
as Fig.~\ref{fig:headers} shows, there is limited space in the probe
packet headers in which to enable this matching
(only the first 28 octets of
the probe packet are encapsulated in the ICMP {\em Time Exceeded}
response).  
Whithin this space, it is necessary to encode a packet identifier,
and not all fields can be used for this purpose if the packets
are to belong to a single flow.
Also, to avoid
ambiguity when there are multiple instances of the program running on
the same machine, both traceroute and \onelab{}  encode a process identifier into the probe.

Some header fields, such as the IP Version, simply cannot be altered
from their original purpose, as routers would tend to discard the
packets as malformed.
Other fields, such as the TTL, Source Address,
and Destination Address, 
naturally cannot be altered.
 Nor would it be suitable to encode identifiers into the IP
options, as packets with IP options are typically processed off the
routers' `fast path'~\cite{govindan2002estimating}, meaning by
different code, and thus with possibly different forwarding rules than
the normal packets whose paths traceroute is supposed to trace.
Finally, as we have mentioned, the IP TOS field and the first four
octets of the transport layer header are off bounds as they are used
for load balancing.

\onelab{} uses the seventh and eight octets of the transport layer
header as the packet identifier.  In UDP probes, this is the Checksum
field.  However, simply setting the checksum value without regard to
packet contents would create ill-formed probe packets liable to be discarded
as corrupt.  To obtain the desired checksum, \onelab{} manipulates the
UDP payload.  

For ICMP {\em Echo} probes, \onelab{} uses the Sequence Number field,
as does classic traceroute.  However, in classic traceroute this has
the effect of varying the ICMP Checksum field, which
is used for load balancing.  \onelab{} maintains a
constant ICMP Checksum by changing the ICMP Identifier field to offset
the change in the Sequence Number field.

\onelab{} also sends TCP probes, unlike classic traceroute, but like
Toren's variant tcptraceroute~\cite{toren2001tcptraceroute}.  To
identify TCP probes, \onelab{} uses the Sequence Number field (instead
of the IP Identification field used by tcptraceroute).  No other
manipulations are necessary in order to maintain the first four octets
of the header field constant.

To encode the process identifier, \onelab{} uses the IP Identification
field, whereas classic traceroute uses the Source Port for UDP probes,
and the Identifier for ICMP Echo probes.  
Tcptraceroute does not encode a process identifier into its probe
packets. 

The simple fact of maintaining a constant five-tuple is not original
to \onelab{}, as tcptraceroute already does this for the TCP probes
that it sends:
in order to more easily traverse firewalls, tcptraceroute by default
sets probes' Destination Port field to 80, emulating web traffic.
No prior work has, however, examined the effect of this choice with
respect to load balancing.

In addition to fixing the flow identifier problem,
\onelab{} provides additional information 
useful for recognizing
certain other traceroute measurement artifacts. 
The \textit{probe TTL} is the TTL that is found in the IP header
of the probe packet encapsulated in the ICMP {\it Time Exceeded}
response.
This value corresponds to the probe's TTL when the router received it and
decided to discard it.
Under normal traceroute behavior, this value is one.
A value other than one reveals an error.  
The \textit{response TTL} is the TTL of the {\it Time Exceeded} response itself.
This value, available also through classic traceroute,
helps in inferring the length of the return path. 
Finally, the \textit{IP ID} is the Identification field from the IP header
of the {\it Time Exceeded} response.
This field is set by the router with the value of an internal 16-bit
counter that is usually incremented for each packet sent. 
The IP ID can help
identify the multiple interfaces of a single router, as described in the
\textit{Rocketfuel} work~\cite{spring2002measuring}, or uncover different
routers and hosts hidden behind a firewall or a NAT box, as described by Bellovin~\cite{bellovin2002nat}.

\medskip
Finally, \onelab{} may use different strategies for sending probes.
\textit{PacketbyPacket} behaves like the classic
traceroute, i.e., sends a probe, waits for an answer or a timeout, 
sends the following probe, and so on. \textit{HopByHop} sends all the 
probes for a given hop with a configurable delay between probes
(the default value of this delay is 50 ms),
waits for answers or timeouts, and then repeats the same procedure for 
the next hop. HopByHop is faster than PacketByPacket.
\textit{Concurrent} sends all the probes for all the hops
with a configurable delay (default 50 ms) between
probes. This algorithm is significantly faster than the previous
two.
To use Concurrent, however, one must know the number of hops needed to reach
the destination.
\textit{Scout} sends one probe with a very high TTL to the
destination. If the destination responds, then it estimates the number of
hops needed to reach the destination and uses Concurrent. Otherwise,
this strategy cannot be used.
Scout is similar to the \textit{raceroute} technique proposed by
Moors~\cite{moors2004streamlining}.

\onelab{} also allows the user to customize the
minimum and maximum TTL values,
in order to focus the measurement on a part of the path,
and thus further speed up the measurements.

\def\rounds{1,465}
\def\duration{74}
\def\probessent{$241$ million}
\def\starinroute{$7.7$ million}
\def\fakevol{$57$ thousand}
\def\fakes{11}

\section{Measurement setup}
\label{measurement_setup}

This section explains how we conducted side-by-side measurements with
classic traceroute and \onelab{}, in order to study traceroute 
measurement artifacts.

Our measurement source was located at
the LIP6 laboratory of the Universit\'{e} Pierre et Marie Curie, in
Paris, France. The university has only one connection to the internet
via the French academic backbone, Renater. 

Our destination list consisted of 5,000  IPv4
addresses chosen at random, without duplicates, that answered to a ping probe at
the time of the creation of the list.
Following the lead of Xia et al.~\cite{xia2005}, we only
considered pingable addresses so as to avoid the artificial inflation
of traceroute measurement artifacts in our results that would come from tracing
towards unused IP addresses.

For a period of $\duration$ days between June and August 2006,
we performed \rounds{} rounds of measurements by using classic traceroute and \onelab{}
to probe from our source to all destinations.
This yielded the data set we use throughout this paper.

This data set serves as a case study, to show
how one can identify, explain, and remove traceroute measurement
artifacts with \onelab{}.  It is not intended to be representative of
the statistics of traceroute measurement artifacts on the internet in
general.
We conducted several other measurements
with similar setups and obtained results consistent with the ones we present below.
Our results should therefore be considered as representative
of what we observe {\em from this source and under these measurement
conditions},
but statistics obtained from other vantage points,
or towards other destination sets, may vary.
Obtaining a representative view of what
can be seen on average on the internet is clearly an interesting
question, but it is out of the scope of this paper.

Each round of measurement was conducted in the following way.
We launched 32  processes in parallel, that each probed $1/32^{nd}$ of the
destination list.
Each process selected, in turn, each destination $d$ from its portion of
the list, and traced two routes to $d$, first using \onelab{},
then using an instance of classic traceroute (NetBSD version 1.4a5).
For both \onelab{} and classic traceroute, we sent a single UDP probe for each hop,
using the PacketByPacket strategy: 
waiting for a reply or a timeout from hop $h$ before sending the
probe for hop $h+1$.\,\footnote{This is the only possible strategy for
classic traceroute.}
The chosen  timeout was $2$ seconds.
\onelab{} kept the five-tuple of its probes constant during each instance
of probing to a given destination, and selected Source and Destination
Port values randomly from the range [10,000, 60,000].
Concerning classic traceroute, we kept the default behavior.
The probing towards a given destination
terminated when either the destination
responded, or an ICMP message other than {\em Time Exceeded} was received.
Moreover, \onelab{} also stopped when TTL 36 was reached,
or when eight consecutive stars were seen, whichever came first.
Classic traceroute was set to stop if it reached a TTL of three greater than the last hop
at which the previous run of \onelab{} received an answer.
Finally, we always set the minimal TTL to $2$ to skip the routers inside
the university network.

One round of measurements to all destinations took approximately one 
hour and fifteen minutes, 
at the rate of approximately 27.3 seconds for both a \onelab{}
and a classic traceroute to a given destination.

For the \probessent{} responses that contain valid IP Source Address
values, we mapped the address to an AS number using Mao et  al.'s
technique~\cite{mao:04}.  Our data set covers 1,498 different ASes,
which corresponds to six percent of the ASes in the internet today.
Our data set covers all nine tier-1 ISP networks and  75 of the one
hundred top-20 ASes of each region according to APNIC's  weekly
routing table report.\,\footnote{APNIC automatically generates  reports
describing the state of internet routing tables. It ranks  ASes for
each of five world regions according to the number of networks
announced.}  Stars mostly appeared at the end of \opphroute{}s
(when a destination does not answer, our measurement method induces
many stars at the end of the corresponding \opphroute{}), with just
\starinroute{} appearing in the midst of responses. We searched
our measurements for invalid IP addresses, i.e., addresses that should
not be given to any host on the public internet~\cite{special-rfc3330}.
We found $\fakes$ invalid IP addresses that account for \fakevol{} 
responses. None of these invalid addresses appear in the structures
we study in the next section, and therefore they  probably play little role in our
observations.  

\medskip
Finally,
we define a \textit{\opphroute{}} to be the output of a given
classic traceroute or \onelab{} instance.
Formally, a \opphroute{} is an
$\ell$-tuple $\mathbf{r}=({\h}_0,\cdots,{\h}_{\ell})$ where ${\h}_0$
is the source address, and, for each $i$, $1 \leqslant i \leqslant \ell, {\h}_i$ stands 
either for the IP address received when probing with TTL $i$, or for
a star if none was received.
The integer $\ell$ is called the \textit{length} of the \opphroute{}.
We call any tuple of the form $(\h_i, \h_{i+1}, \dots, \h_{i+k})$
a \textit{sub\-route} of $\mathbf{r}$ of length $k$.

\section{Structures under study}\label{anomalies}

We focus our observations on three particular structures that 
appear in many \opphroute{}s: we call them loops, cycles, and diamonds.
In this section we describe these structures,
and present some basic statistics about their frequency of appearance
in our traceroute data set.
Sec.~\ref{loops} discusses loops, Sec.~\ref{cycles} discusses
cycles, and Sec.~\ref{diamonds} discusses diamonds.
In subsequent sections,
we give possible causes and explanations for these structures,
as well as methods for distinguishing between the different causes.

\def\nodefreqloops{7.51}
\def\destfreqloops{24.6}
\def\routefreqloops{4.35}
\def\persistentloopvolume{7.45}

\subsection{Loops}\label{loops}
In some \opphroute{}s, the same IP address appears
twice or more in a row: we call this a \textit{loop}.
In the normal course of routing, a router does not forward a packet 
back to the incoming interface. Hence, loops are
most likely an artifact of the measurement itself.
Formally, a loop is observed on IP address $\h_i$ with destination $d$ if there is 
at least one \opphroute{} towards $d$ containing $ \cdots , \h_i, \h_{i+1}, \cdots$ with
$\h_{i+1} = \h_i$.
The term `IP address' implies that $\h_i$ is not a star.

\begin{figure}[!ht]
\begin{center}
\includegraphics[scale=0.4]{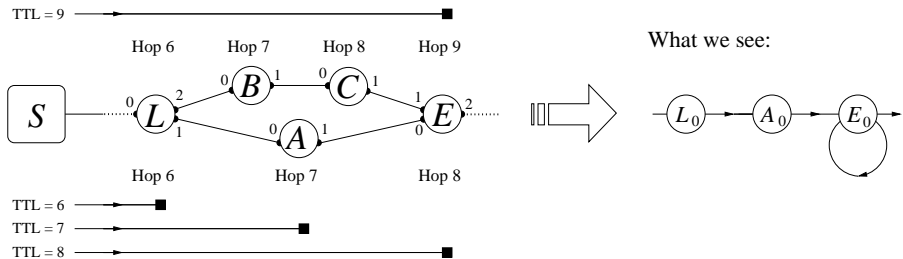}
\end{center}
\caption{Loop caused by load balancing.}
\label{fig:loop-loadbalancing}
\vspace{-3mm}\end{figure}

Load balancing can cause loops that appear
sporadically when repeatedly tracing from a source to the
same destination;
this is shown in
Fig.~\ref{fig:loop-loadbalancing}. These loops can occur in the presence of
a load balanced route with at least two paths having a length difference
equal to one.

We call each observation of a \opphroute{} as described above
a {\em loop instance}, and
we define this loop's {\em signature} as the pair $(\h_i,d)$.
For instance, if we have four \opphroute{}s $(\ldots,a,a,b,\ldots,d_1)$,
$(\ldots,a,a,b,\ldots,d_2)$, $(\ldots,a,a,b,$ $\ldots,d_2)$ and
$(\ldots,a,a,c,\ldots,d_2)$, then we have four loop instances (one for
each \opphroute{}), and two loops signatures ($(a,d_1)$ and $(a,d_2)$).
Depending on the context, we will focus on
statistics involving either loop signatures or loop instances. However,
when the context is clear or when such distinction is irrelevant, we
will simply use the term ``loop''.

Moreover, we say that a \opphroute{} $\mathbf{r}$ contains a $n$-loop instance,
with $n \geqslant 1$, on IP address $\h$ when $\h$ was observed on exactly $n+1$
consecutive positions on $\mathbf{r}$
($n$ is the number of times $\h$ is observed at two consecutive positions).
We call $n$ the $length$ of the loop instance. 
The shortest loops are of length 1.
We extend this
definition to loop signatures: the length of a loop signature is the 
maximal length of all its instances.

Note that we use the term ``loop'' as meant commonly in graph theory.
This is different from a forwarding loop, which typically
involves several addresses.  Forwarding loops are best described as
cycles, which are discussed in Sec.~\ref{cycles}.

Using the above definitions, enumerating loops
is straightforward: we simply scan all the \opphroute{}s and look
for instances of \ip{} addresses that appear at least twice consecutively.
To perform statistics, we maintain a list of all loop signatures encountered,
and for each signature $(\h,d)$ we keep a record of the relevant information:
the number of instances, the maximal length, the number of times
the \ip{} address $\h$ has been observed on \opphroute{}s towards $d$
(whether a loop has been observed or not), and so on.

\bigskip
Loops appear to be surprisingly common:
overall, $\nodefreqloops \%$ of the IP addresses detected in our experiment
were in at least one loop instance, and more than
$\routefreqloops \%$ of the \opphroute{}s contained a loop.
Moreover, when probing repeatedly, say $N$ times to the same destination, 
the probability that at least one of the \opphroute{}s contained a loop
increased with time; during our measurements, we were able to
observe loops towards more than $\destfreqloops \%$ of the destinations.
This is due to the high heterogeneity of loops: while some of them seem to
be {\em persistent} and appeared on almost every \opphroute{} towards 
their destination, many others are {\em occasional} and appeared only once, 
or a couple of times.
Measuring for a longer period of time therefore increased the probability of
observing such occasional loops, within the duration of our experiments.\,\footnote{%
This probability would, however, most likely stop growing given sufficiently long
measurements, as we cannot expect to observe loops on \opphroute{}s to every single destination.}
We also observed an intermediate behavior: {\em systematic} loops.
We say that a loop $(\h,d)$ is systematic if, whenever the \ip{} address $\h$
appeared in a \opphroute{} towards $d$, this occurrence was part of a loop.
In other words, every time we saw $\h$,
we saw the loop.

The difference between systematic and persistent loops is that systematic
loops don't always show up on \opphroute{}s to a given destination, and may actually be exceptional.
To examine this further, we defined two distinct characteristics. Consider a loop $(\h,d)$
on address $\h$ on a \opphroute{} towards a destination $d$:
\begin{enumerate}
\item Its \emph{appearance frequency} is the probability that
 a \opphroute{} towards $d$ contained the loop.
\item Its \emph{conditional appearance frequency} is the probability
 that a \opphroute{} towards $d$ \emph{that contained $\h$} also contained
 the loop.
\end{enumerate}

\begin{figure}[!ht]
\begin{center}
\includegraphics[scale=0.4]{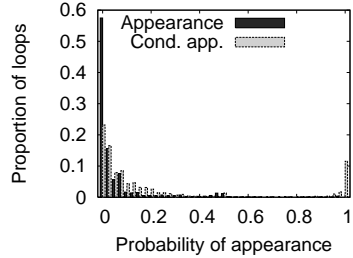}
\end{center}
\caption{Dark bars: appearance frequency of the loops.
Light bars: conditional appearance frequency of the loops.}
\label{fig:loop-appearance}
\end{figure}

Fig.~\ref{fig:loop-appearance} shows the distribution of these two
characteristics over the loop signatures we observed.
Notice that the proportion of
persistent loops (appearance frequency close to $1$) seems to be very small,
much smaller than the proportion of systematic loops (conditional appearance
frequency close to $1$). However, since these loops were observed more often
than the others, their proportion as \emph{instances} was much more significant
($\persistentloopvolume \%$ of all instances).

\begin{figure}[!ht]
\begin{center}
\includegraphics[scale=0.4]{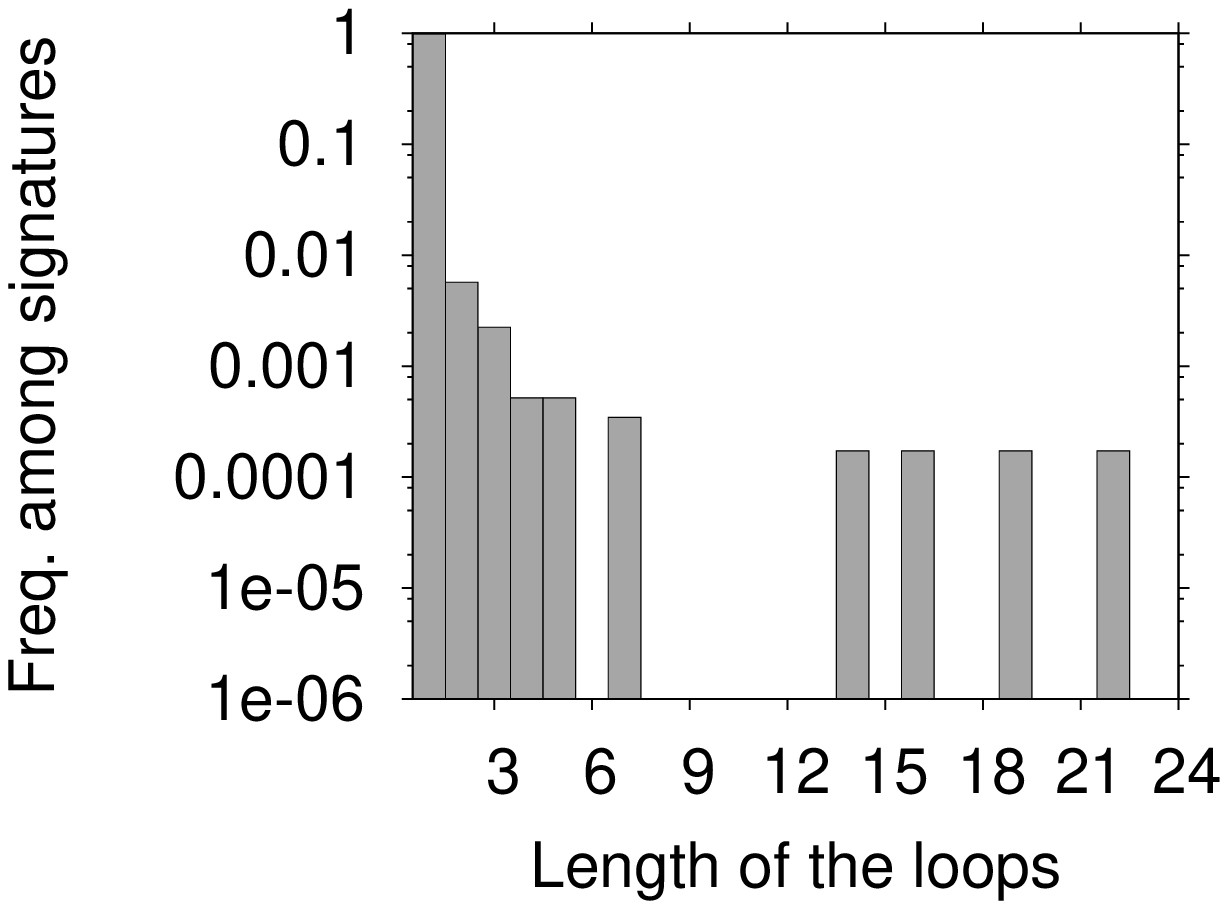}
\includegraphics[scale=0.4]{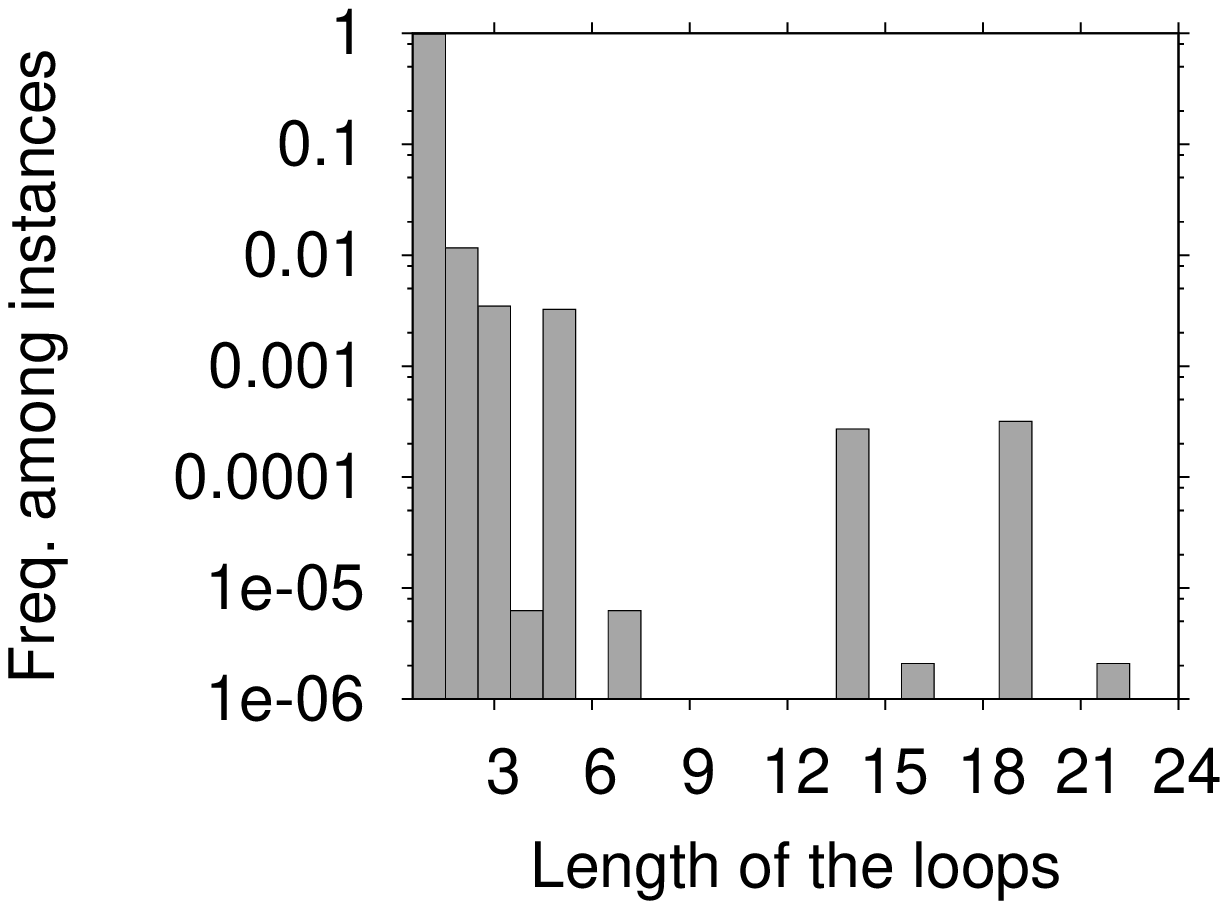}
\end{center}
\caption{Distribution of the length of loop signatures (left) and instances
 (right). 
}
\label{fig:loop-length}
\end{figure}

Fig.~\ref{fig:loop-length} shows (left) the {\em lengths}
of the \nloops\ we observed and their distributions among loop
signatures (left) and loop instances (right).
This chart is logarithmic, since 
large \nloops\ (i.e., with $n>1$) are very uncommon compared to $1$-loops.
However, we observed several persistent large \nloops{}, which indicates
that they shouldn't be considered as marginal events. 

Both length and appearance statistics confirm that several `flavors' of loops
exist. This points to the unlikelihood that all loops are generated by
a unique mechanism, and suggests characterization
of the loops into different categories, each of them resulting from
a different cause.

\def\nodefreqcycles{4.73}
\def\destfreqcycles{17.5}
\def\routefreqcycles{0.60}

\def\cyclesignaturesclassic{5,674}
\def\persistentcycleratiovol{3.93}
\def\systematiccyleratiovol{8.95}

\subsection{Cycles}
\label{cycles}

Formally, a \opphroute{} $\mathbf{r}$ is said to be \emph{cyclic} on an IP address $\h$,
or $\h$-cyclic,
if it contains $\h$ at least twice, at {\em nonconsecutive} locations,
i.e., separated by at least one IP address $\h'$ distinct from $\h$.
The term `IP address' implies that neither  $\h$ nor $\h'$ are stars.
This distinction is to make sure we don't misinterpret possible $n$-loops
as cycles.
For instance, a \opphroute{} $(\ldots, b,c,d,e,c,f, \ldots)$ is cyclic on $c$,
but the \opphroute{} $(\ldots, b,c,*,*,c,f, \ldots)$ 
is not, since the sequence $c,*,*,c$ might well be a $3$-loop.

Load balancing can cause cycles, in the same way as loops; see Sec.~\ref{loops} and
Fig.~\ref{fig:loop-loadbalancing}.
Cycles can occur when there is load balancing
between routes having
a length difference larger than one.

As for loops, we use the term {\em cycle instance} for any occurrence of a cycle
on a \opphroute{}, and define a \emph{cycle signature} as a pair $(\h, d)$ of
an IP address and a destination such that at least one of the
\opphroute{}s towards $d$ is cyclic on $\h$.
Most of the cycles we observed occurred at more than one location on the \opphroute{}s,
and sometimes they even overlap with each other,
making it hard to define properties of a cycle as if it
were a single, well-defined object.
To be able to provide some statistics, we consider two properties.
The \emph{length} of a cycle signature $(\h,d)$ is the {\em shortest} distance
that separates two instances of $\h$ in all \opphroute{}s towards $d$ that
are $\h$-cyclic.
The \emph{span} of a cycle signature $(\h,d)$ is the 
{\em greatest} distance that separates two instances of $\h$ in 
all \opphroute{}s towards $d$ that are $\h$-cyclic.
These properties 
capture respectively the minimal and maximal length of a round-trip 
from $\h$ to $\h$ observed on a \opphroute{} towards destination $d$.
For instance, if we have three \opphroute{}s $(\ldots, r,a,b,r,\ldots, d_1)$,
$(\ldots, r,a,b,r,\ldots, d_2)$ and $(\ldots,r,c,d,e,r,\ldots, d_2)$,
then the cycle signature $(r,d_1)$ has
both length and span equal to $3$,
whereas the cycle signature $(r,d_2)$ has a length of $3$ and a span of $4$.

An interesting species of cycles is the \emph{periodic cycles},
which are encountered in \opphroute{}s like $(\ldots, a,b,c,b,c, \ldots)$.
A \opphroute{} is \emph{$k$-periodically cyclic} on IP addresses
$\h_1, \h_2, \cdots, \h_k$ when this sequence of IP addresses appear
at least twice, consecutively, and in the same order, on the \opphroute{}.
Formally, a periodic cycle signature is a pair $((\h_1,\cdots,\h_k), d)$ such
that at least one \opphroute{} towards $d$ is $k$-periodically cyclic
on $\h_1, \cdots, \h_k$.

As for loops, enumerating cycles consists in
scanning every \opphroute{}, detecting the cycle instances,
and aggregating information about every cycle signature encountered.

\bigskip
Cycles are less common than loops in our data set: they appeared on only
$\routefreqcycles \%$ of the \opphroute{}s.
On the other hand, they appeared on a broad range of addresses:
we observed cyclic \opphroute{}s towards
$\destfreqcycles \%$
of the destinations, and $\nodefreqcycles \%$ of the
IP addresses discovered during our experiment
appear in at least one cycle signature.

\begin{figure}[!ht]
\begin{center}
\includegraphics[scale=0.4]{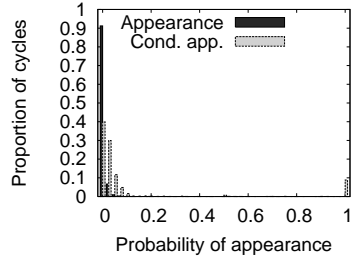}
\end{center}
\caption{Dark bars: appearance frequency of cycles. Light bars:
conditional appearance frequency of the cycles.}
\label{fig:cycle-appearance}
\end{figure}

Fig.~\ref{fig:cycle-appearance} presents appearance statistics
similar to the ones already described
for loops, using a similar terminology.
To avoid redundancies, we invite the reader to consult Sec.~\ref{loops}
for definitions.
Persistent cycles appear to be very rare; they are actually invisible on
Fig.~\ref{fig:cycle-appearance}. We did detect $2$ persistent cycles over
the \cyclesignaturesclassic{} cycle signatures we observed, and they 
account for $\persistentcycleratiovol \%$ of the cycle instances.
Systematic cycles were more common (see the peak in
    Fig.~\ref{fig:cycle-appearance} for a conditional appearance
    frequency of $1$), and represented $\systematiccyleratiovol \%$ of the
cycle instances.  They still have a very low appearance frequency, meaning
that in spite of their systematic behavior, their occurrence remained infrequent
in our measurements.
Overall, it seems that cycles, unlike loops, are mostly \emph{occasional}
events.
This makes them more difficult to track.

\begin{figure}[!ht]
\begin{center}
\includegraphics[scale=0.4]{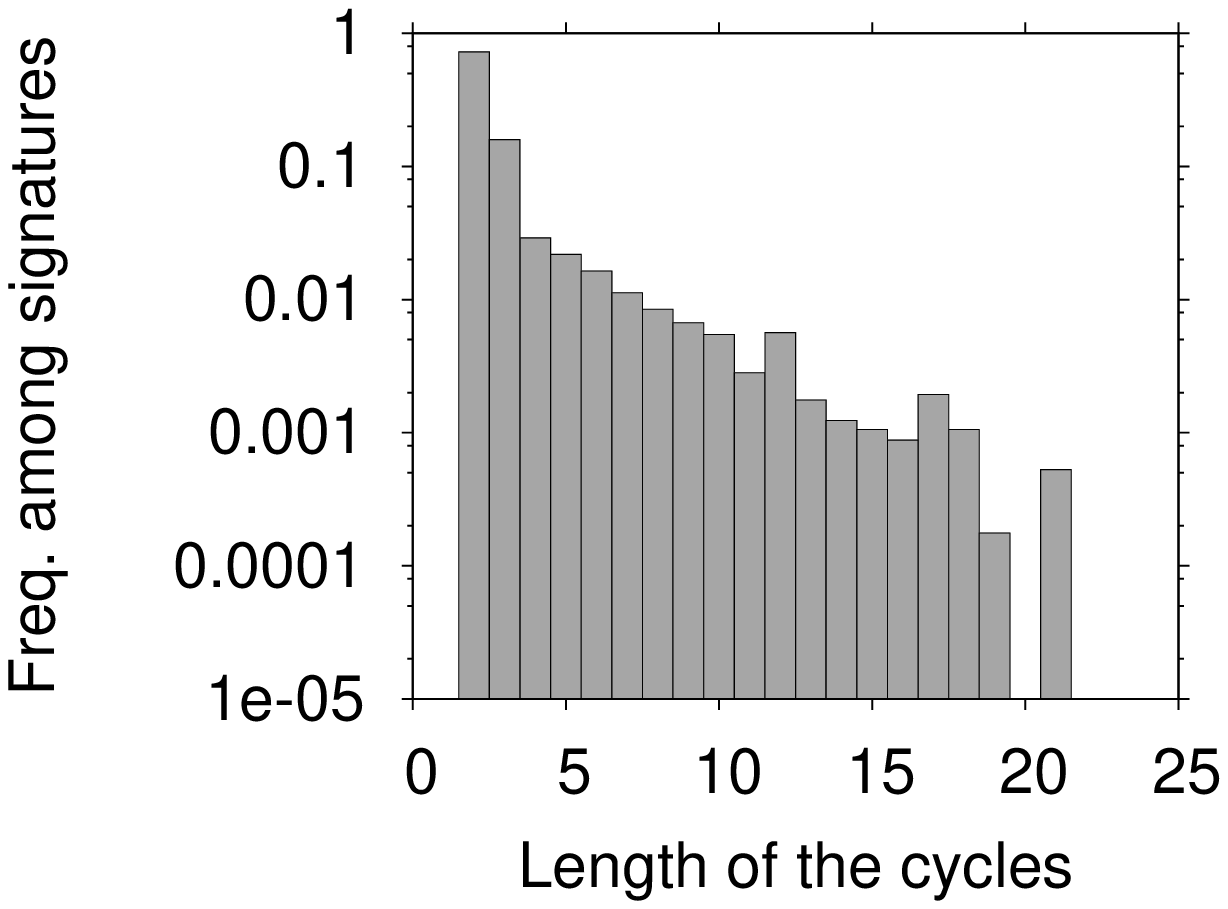}
\includegraphics[scale=0.4]{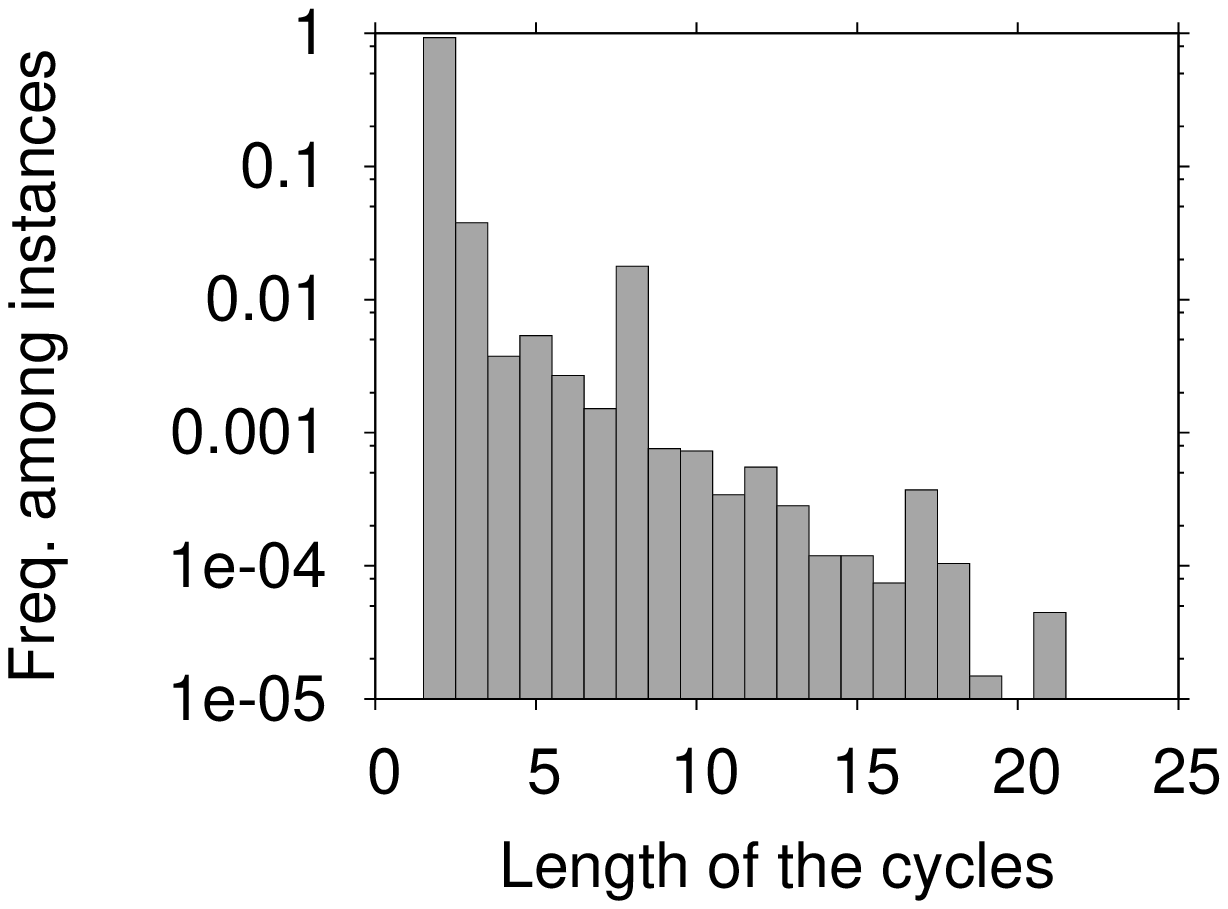}

\vspace{0.3cm}

\includegraphics[scale=0.4]{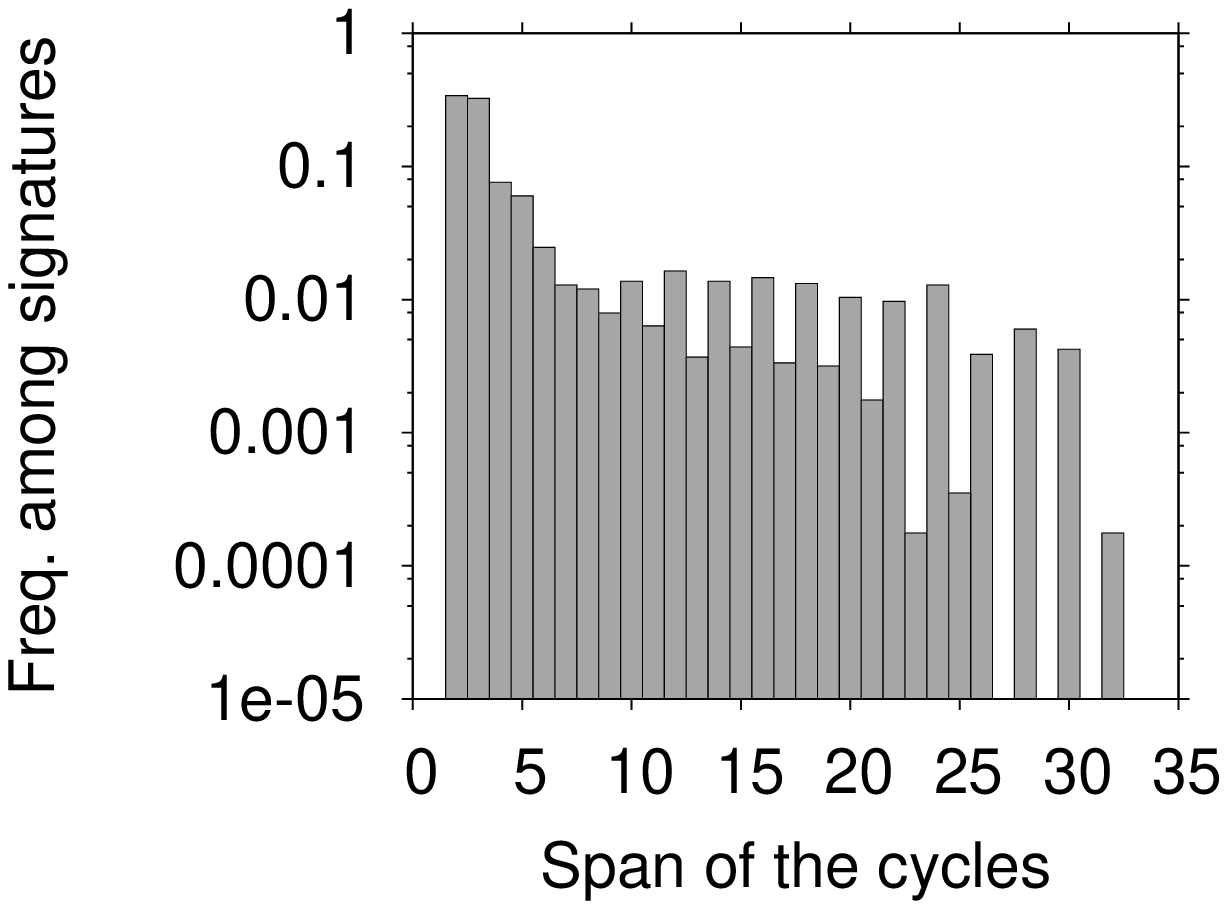}
\includegraphics[scale=0.4]{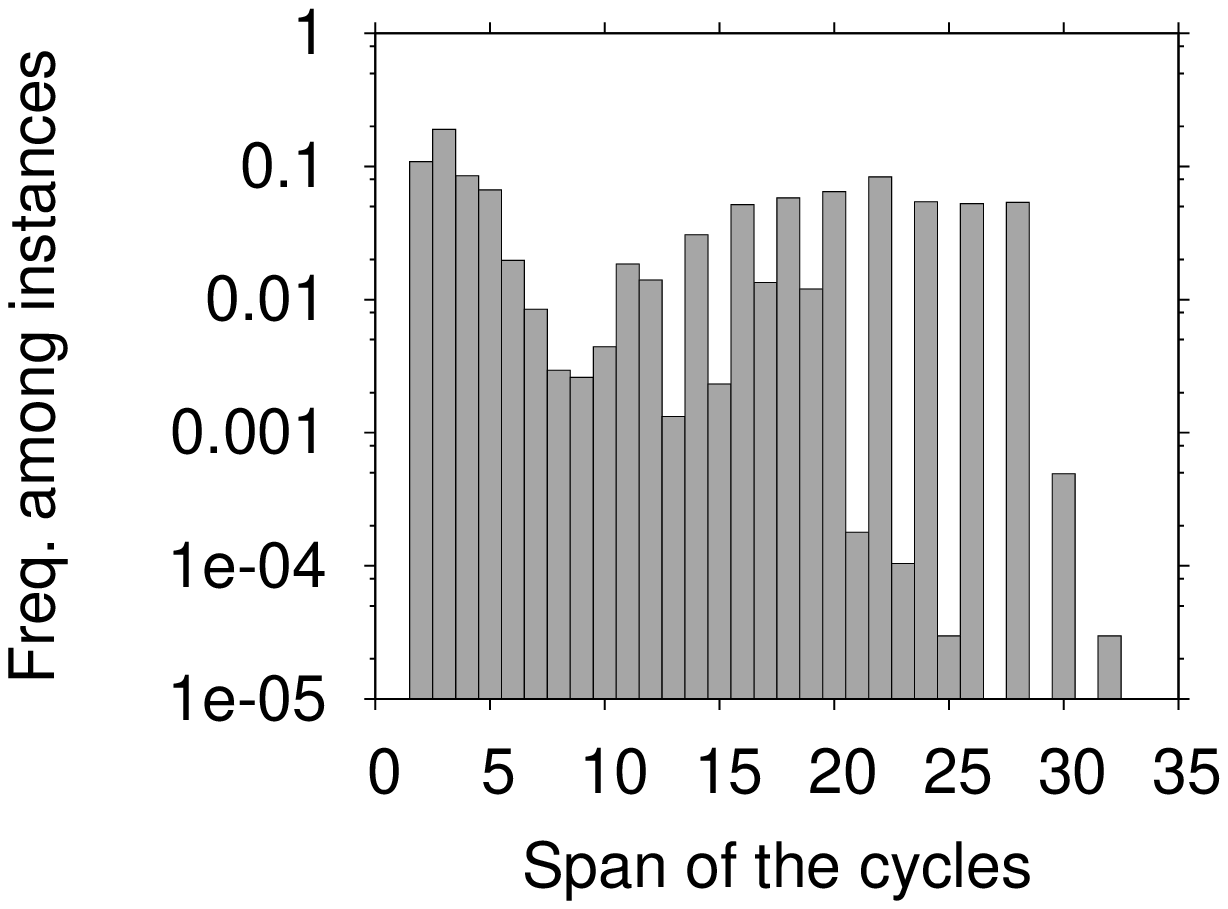}
\end{center}
\vspace{-0.3cm}
\caption{Distribution of the length (top) and span (bottom) of the cycles
signatures (left) and instances (right).}
\label{fig:cycle-length-span}
\end{figure}

Fig.~\ref{fig:cycle-length-span} shows the 
distribution of the length and span of the cycles we observed (as for loops,
    we both use the signature- and instance-based statistics). 
These statistics illustrate the
variety of cycles that can be observed.

From the distribution of the length of cycles, it is clear that cycles of
length $2$ were predominant, especially for cycle
instances. However, the cycles having a span of $2$ were not as common,
and they were even outnumbered by cycles of span $3$ in terms of instances,
which indicates that not all cycles of length $2$ also have a span equal to $2$.
This is related to the observation that, for higher span values,
cycles with an even span are much more likely
to occur than those with odd span. These observations can be explained in terms of the
periodic cycles we introduced earlier: a periodic cycle of period $2$
will have a length equal to $2$, and may have any even span between $2$
and the maximum number of hops allowed in our measurements.

This, plus the fact that cycles of extreme length also occurred regularly
(according to the length distribution), also suggests the existence of
a wide variety of phenomena as possible causes for cycles.

\subsection{Diamonds}
\label{diamonds}

\newcommand{\onedstdi}{one-destination diamond}
\newcommand{\Onedstdi}{One-destination diamond}
\newcommand{\globaldi}{global diamond}
\newcommand{\Globaldi}{Global diamond}
\newcommand{\ddi}[1][d]{\ensuremath{#1}-diamond}

Loops and cycles are structures observed on single \opphroute{}s.
Other types of structures in traceroute measurements appear only when
multiple \opphroute{}s are considered together
(e.g., when constructing maps of the
network).
A typical feature that we observe in this way is what we call a diamond.

Given a set $S$ of \opphroute{}s, a \textit{diamond} 
is a pair $(h,t)$ of IP addresses for which the number $k$ of distinct
IP addresses ${\h}_i$ such that there exists a \opphroute{} in $S$ 
of the form $\ldots,h,{\h}_i,t,\ldots$ is at least $2$.
The term `IP address' implies that neither $h$, $t$, nor any of the $\h_i$ are stars.
 We call $h$ the
 \textit{head} of the diamond, and $t$ its \textit{tail}.  
The \textit{core} of the diamond is the set of addresses
$\{{\h}_1,\ldots, {\h}_k$\},
and the diamond's {\em size} is $k$.

\begin{figure}[ht!]
\centering
\includegraphics[scale=.4]{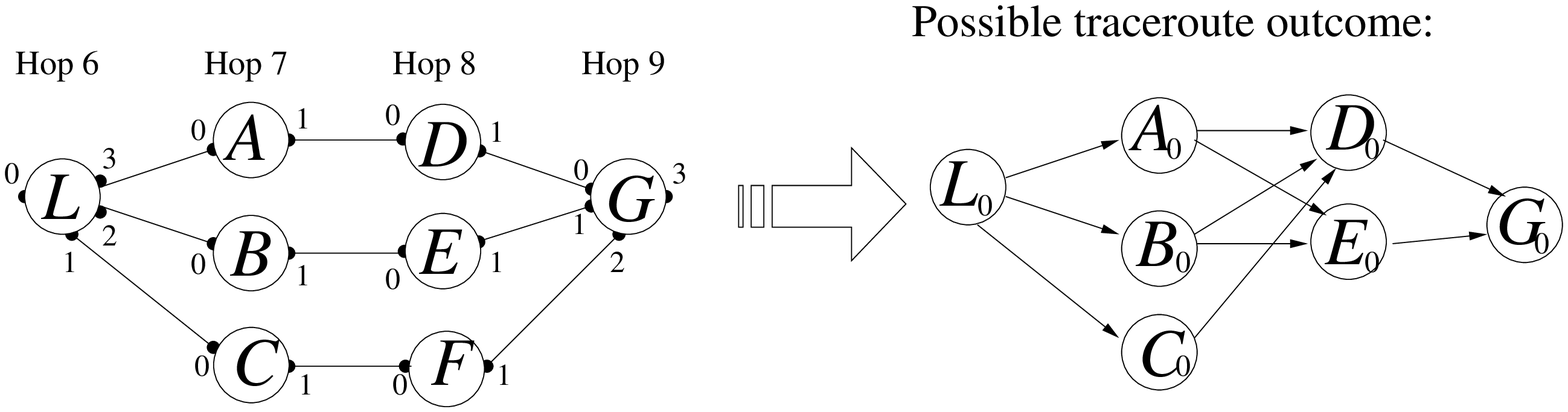}
\caption{An example of several diamonds caused by load balancing.
For clarity, we omit the probe packets.}
\label{fig_ex_diamonds}
\end{figure}

Load balancing may induce many diamonds.
Fig.~\ref{fig_ex_diamonds} presents a typical such case.

Which diamonds we see depends upon which set $S$ of \opphroute{}s is
considered.  We will see in Sec.~\ref{impact_di} that the distinctions 
between different types of diamonds afford interesting observations
concerning load balancing.

A diamond $(h,t)$ that emerges from
only the \opphroute{}s towards a single destination $d$ is called
 a {\em \ddi{}}.
If there exists at least one
destination $d$ for which $(h,t)$ is a \ddi{}, then we call $(h,t)$ a {\em
  \onedstdi{}}.  We define the size of a \onedstdi{} $(h,t)$ to be the
maximum of the sizes of all \ddi{}s $(h,t)$.

A diamond $(h,t)$ that emerges from
the entirety of \opphroute{}s in our data set is called  a
{\em \globaldi{}}.  All \onedstdi{}s are also \globaldi{}s.

\begin{figure}[h!]
\begin{center}
\includegraphics[scale=0.4]{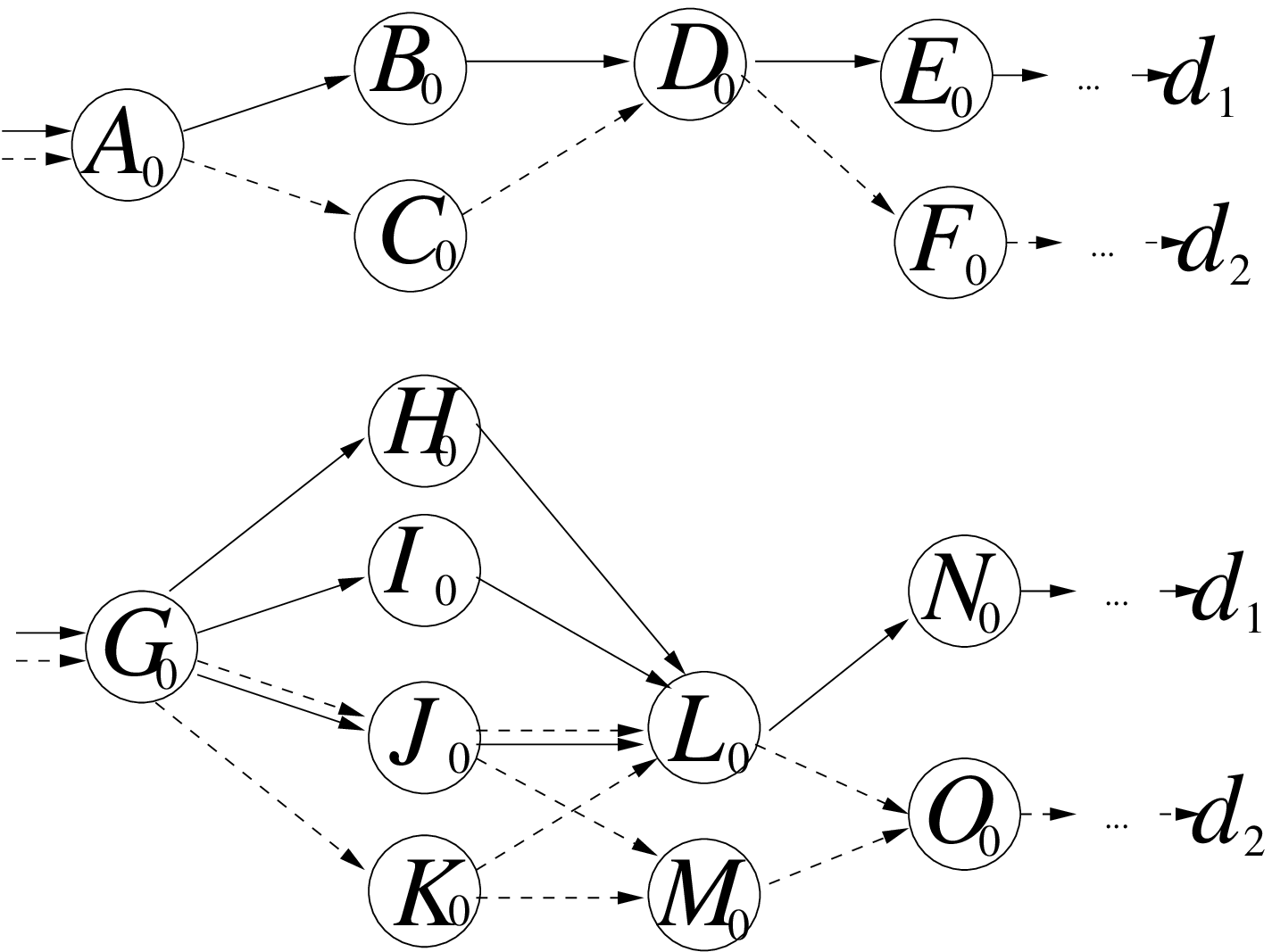}
\end{center}
\caption{Examples of \globaldi{}s, \ddi{}s, and \onedstdi{}s.}
\label{fig_ex_types_diamonds}
\end{figure}

Fig.~\ref{fig_ex_types_diamonds} presents examples of \ddi{}s, \onedstdi{}s,
and \globaldi{}s.
We see \opphroute{}s towards two distinct destinations
$d_1$ and $d_2$.
$A_0, \ldots, O_0$ are addresses observed on these \opphroute{}s,
and a solid (respectively, dashed) arrow between two addresses indicates that they
are linked in a \opphroute{} towards $d_1$ (respectively, $d_2$).
$(G_0, L_0)$ is a \ddi[d_1] of size $3$, and a \ddi[d_2] of size $2$.
Hence $(G_0, L_0)$ is also a \onedstdi{}, and its size is $3$,
which is the size of the largest \ddi{}  $(G_0, L_0)$.
$(G_0, M_0)$, $(J_0, O_0)$ and $(K_0, O_0)$ are \ddi[d_2]{}s of size $2$ and
also \onedstdi{}s of size $2$.
Note that $(A_0, D_0)$ is not a \ddi{} for any $d$, and therefore not a \onedstdi{}.
$(A_0, D_0)$, $(G_0, L_0)$, $(G_0, M_0$), $(J_0, O_0)$ and $(K_0, O_0)$ are \globaldi{}s.

Diamonds are detected in the following manner:
we scan every \opphroute{}, and maintain, for each possible triple $(h,t,d)$
of two \ip{} addresses and a destination,
the set $A_d(h,t)$ of all \ip{} addresses seen
between $h$ and $t$ on \opphroute{}s towards $d$.
We then compute the set $A_{all}(h,t)$ of \ip{} addresses seen
between $h$ and $t$ on {\em all} \opphroute{}s by merging all the $A_d(h,t)$.
For any destination $d$, the \ddi{}s are then the
pairs $(h,t)$ such that $|A_d(h,t)| \geqslant 2$.
\Onedstdi{}s are the pairs $(h,t)$ such that there exists a $d$
such that $|A_d(h,t)| \geqslant 2$, and
\globaldi{}s are the
pairs $(h,t)$ such that $|A_{all}(h,t)| \geqslant 2$.

\begin{figure}[ht!]
\centering
\includegraphics[scale=.4]{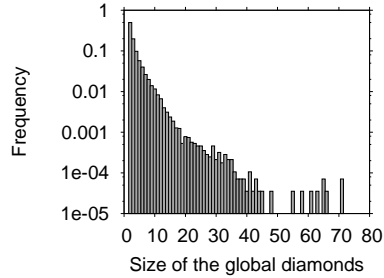}
\caption{Distribution of the size of the \globaldi{}s.}
\label{fig_size_d_tr_multi}
\end{figure}

\bigskip
Overall, we observed 28,231 \globaldi{}s in our traceroute data set.
Fig.~\ref{fig_size_d_tr_multi} presents their size distribution.
They involved a very large fraction of the observed IP addresses:
overall $44.6\%$ of the addresses were involved in \globaldi{}s.
Moreover, \globaldi{}s were not separate entities but were highly interlocked
with each other:
$16.4\%$ of all addresses were in the head of one or more \globaldi{}s,
$30.7\%$ in the core, and $27.1\%$ in the tail.
The sum of these proportions is much larger than $44.6\%$,
which indicates that a large number of addresses belonged to several
\globaldi{}s.

\Onedstdi{}s and \ddi{}s were also quite frequent in our measurements:
there were 95,936 \ddi{}s, and 24,326 \onedstdi{}s.
Among all destinations $d$, $90.2\%$ led to the observation of at
least one \ddi{}.
The average number of distinct \ddi{}s observed with these
destinations was $21.3$.

\begin{figure}[ht!]
\centering
\includegraphics[scale=0.4]{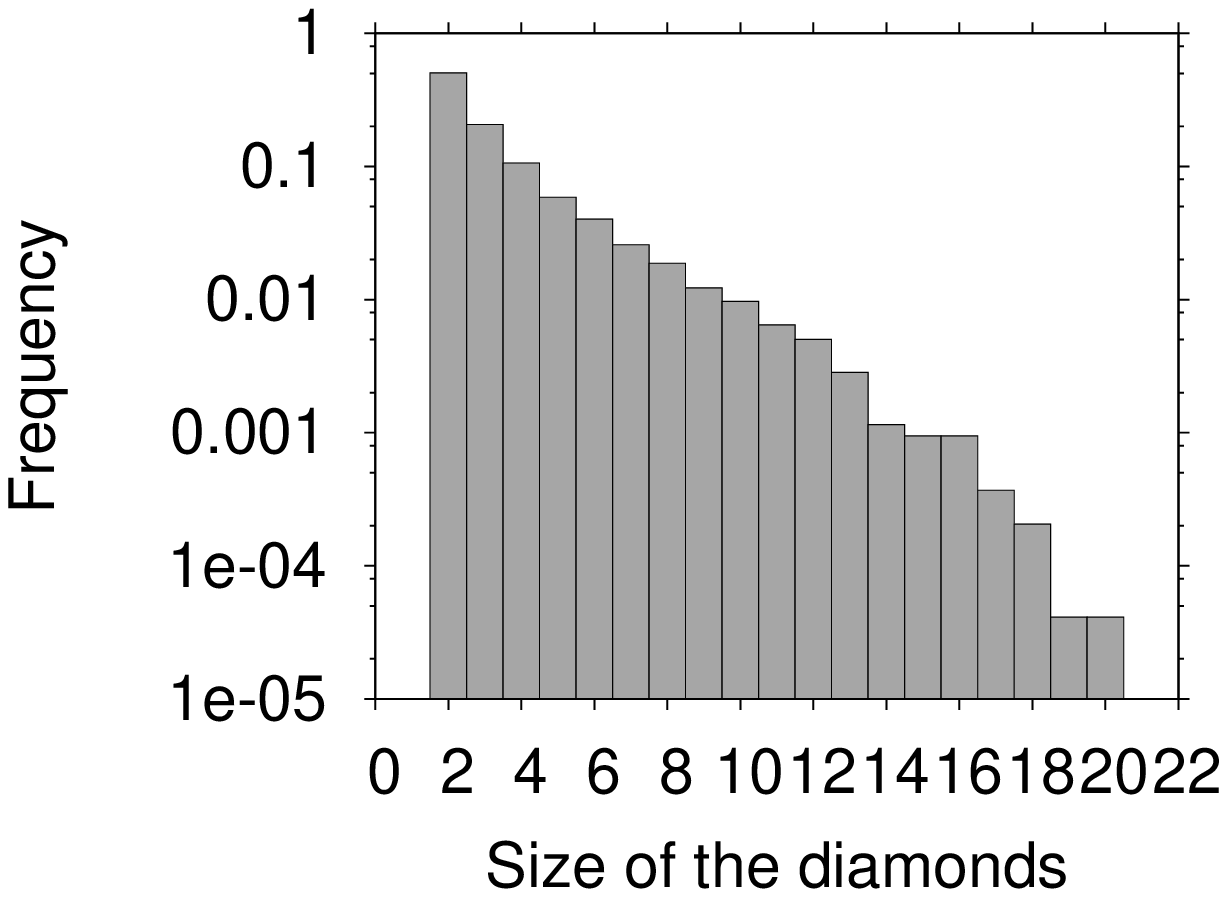}
\includegraphics[scale=0.4]{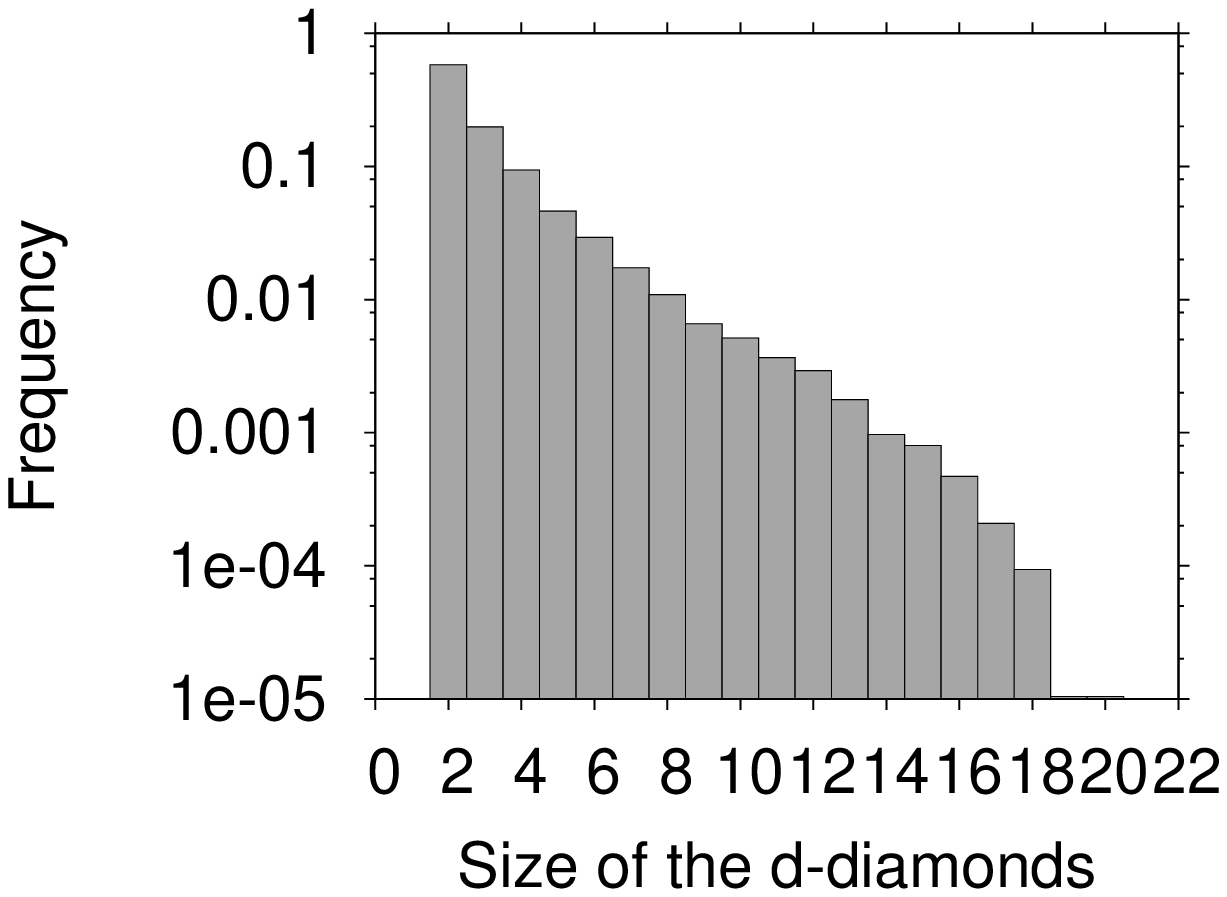}
\caption{Distribution of the size of \onedstdi{}s (left) and \ddi{}s (right).}
\label{fig_size_d_tr_single}
\end{figure}

Fig.~\ref{fig_size_d_tr_single} presents the distribution of the size of \ddi{}s and \onedstdi{}s.
We observe that there were far fewer large \ddi{}s than \globaldi{}s.
This, together with the fact that there are more \globaldi{}s than \onedstdi{}s,
indicates that \opphroute{}s towards different destinations
combined with each other,
creating \globaldi{}s where there were no \onedstdi{}s
(as illustrated in Fig.~\ref{fig_ex_types_diamonds}),
and increasing the size of existing \onedstdi{}s to form large \globaldi{}s.

\def\disappearloop{89.8}
\def\appearloop{2.73}
\def\disappearloopvol{79.9}
\def\appearloopvol{0.21}
\def\disappearcycle{79.5}
\def\appearcycle{11.0}
\def\disappearcyclevol{39.7}
\def\appearcyclevol{1.32}
\def\disappearsystematicloops{89.3}
\def\persistentloopsvol{7.45}
\def\systematiccyclesratiovol{29.0}
\def\persistentcyclesratiovol{3.93}

\section{Measurement artifacts due to load balancing}
\label{controlled_tuples}

We now turn to the study of the differences observed when probing with \onelab{}
rather than with traceroute, for each of the three types of structures we study.

\subsection{Loops}
\label{loop-impact}
We compare the measurements obtained with
classic \traceroute\ and the ones obtained with {\onelab}, the latter
avoiding per-flow load balancing.
Of the loop signatures, $\disappearloop \%$
disappear; however, new loop signatures also appear, which represent
approximately $\appearloop \%$ of the initial total.
For loop instances, the statistics are
$\disappearloopvol \%$ and $\appearloopvol \%$.
This shows that load balancing is likely the primary cause
of loops in our observations.
The fact that some loops are observed in our data set with \onelab{} but
not with classic \traceroute{} most likely comes from the fact that some loops are rarely
observed, as explained in Sec.~\ref{loops}.
It is therefore natural that such loops might only be observed with
one measurement tool and not the other.\,\footnote{%
This also implies that a small fraction of the loops that disappeared when
considering \onelab{} rather than classic \traceroute{} in our data set may have
not been caused by load balancing but by rare events.}
These observations also hold for cycles and diamonds.

\begin{figure}[!ht]
\begin{center}
\includegraphics[scale=0.4]{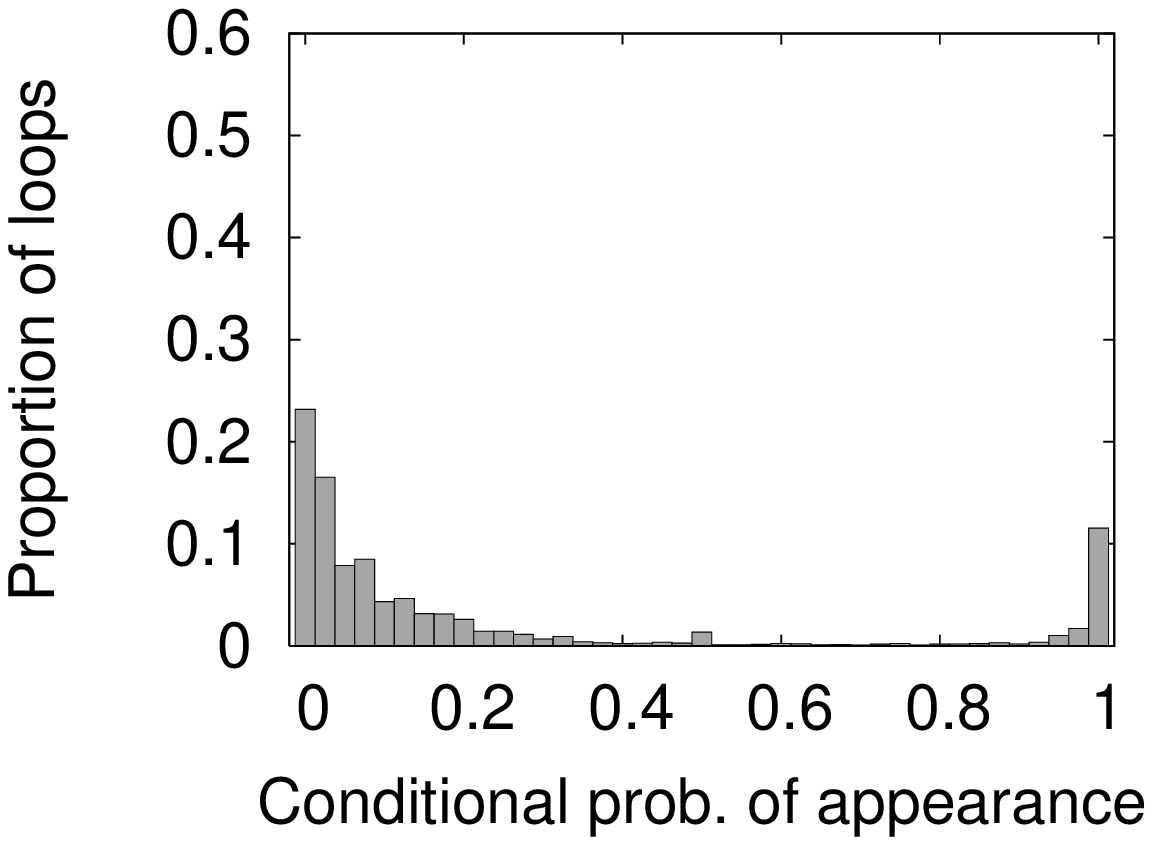}
\includegraphics[scale=0.4]{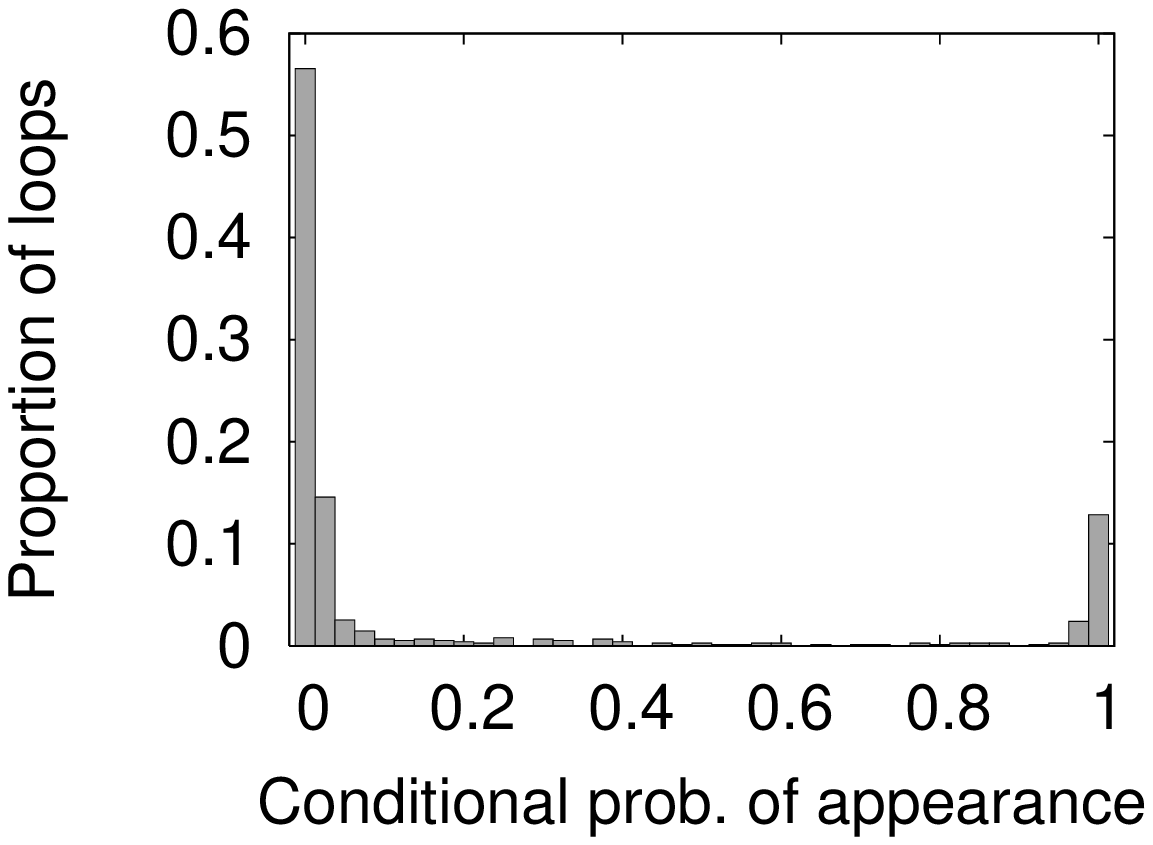}
\end{center}
\caption{Conditional appearance frequencies of loops observed in our 
data set with classic traceroute (left) and  \onelab{} (right).}
\label{fig:loop_difference}
\end{figure}

We investigate this matter further
by looking at the characteristics of the loops removed by
\onelab{}. Fig.~\ref{fig:loop_difference} shows the
conditional appearance frequencies (see Sec.~\ref{loops}) of loops
found in the traceroute and \onelab{} data sets. 
One can see that
almost all loops that were neither systematic nor truly rare, i.e., loops
appearing sporadically, but somewhat regularly -- which is the kind of
behavior one would expect from loops caused by load balancing -- are removed.
At the same time, all persistent loops 
that appeared with traceroute were also present when using \onelab{},
which was expected, since persistent loops are unlikely to be caused
by load balancing.
This will be discussed further in Sec.~\ref{other_causes}.

We also observe that $\disappearsystematicloops \%$
of the non-persistent systematic loops disappear. 
We have seen how a load-balanced route with a pair of path lengths
that differ by 1 can lead to the observation of a loop.  We now
hypothesize a router implementation of per-flow load balancing that
would cause such a loop to be both systematic and non-persistent.  We
use Fig.~\ref{fig:loop-loadbalancing} as an illustration.  If the
router $L$ were to forward probe packets in a round-robin fashion,
alternatively to $B$ and to $A$, then 
our
experiments would reveal only two measured routes out of the many
that are possible: one with the subroute $(L_0,B_0,E_0,E_0)$ and one
with $(L_0,A_0,C_0,F_0)$, where $F_0$ is the next hop beyond $E_0$.
The first subroute contains a loop that is systematic, because it
appears whenever $E_0$ appears, and that is non-persistent, because it
only appears on a portion (in this case, 50\%) of the routes to the
destination.

What would account for such round-robin forwarding at a per-flow load
balancer?  Recall that per-flow load balancing assigns packets to
outgoing interfaces as a function of the 5-tuple, and that classic
traceroute increments the Destination Port by 1 with each subsequent
probe, leaving the rest of the 5-tuple unchanged.  We obtain the
hypothesized behavior if the function employed by the router cycles
through each interface in turn with each increment in the Destination
Port.

These observations confirm that load balancing,
be it flow-based or packet-based,
is an important source of
loops.
However, we will see in Sec.~\ref{other_causes} that
some loops, which represent a significant part of the total
($\persistentloopsvol \%$), are not due to load balancing.
This is true in particular for persistent loops.

\subsection{Cycles}
As for loops, we compare the cycles observed with classic traceroute
and those obtained with \onelab{}.
We observe that $\disappearcycle \%$ of
cycle signatures disappear under \onelab{}, and that new signatures, representing
$\appearcycle \%$ of the initial total, appear.
For cycle instances, these numbers are
$\disappearcyclevol \%$ and $\appearcyclevol \%$.
The diagnosis here still is that load balancing is an important cause, and most
probably the prominent one, for cycles. The 
low number of
cycles 
caused by load balancing 
can be attributed to the unlikelihood of load balancing across paths having
a length difference of two or more.

\begin{figure}[!ht]
\begin{center}
\includegraphics[scale=0.4]{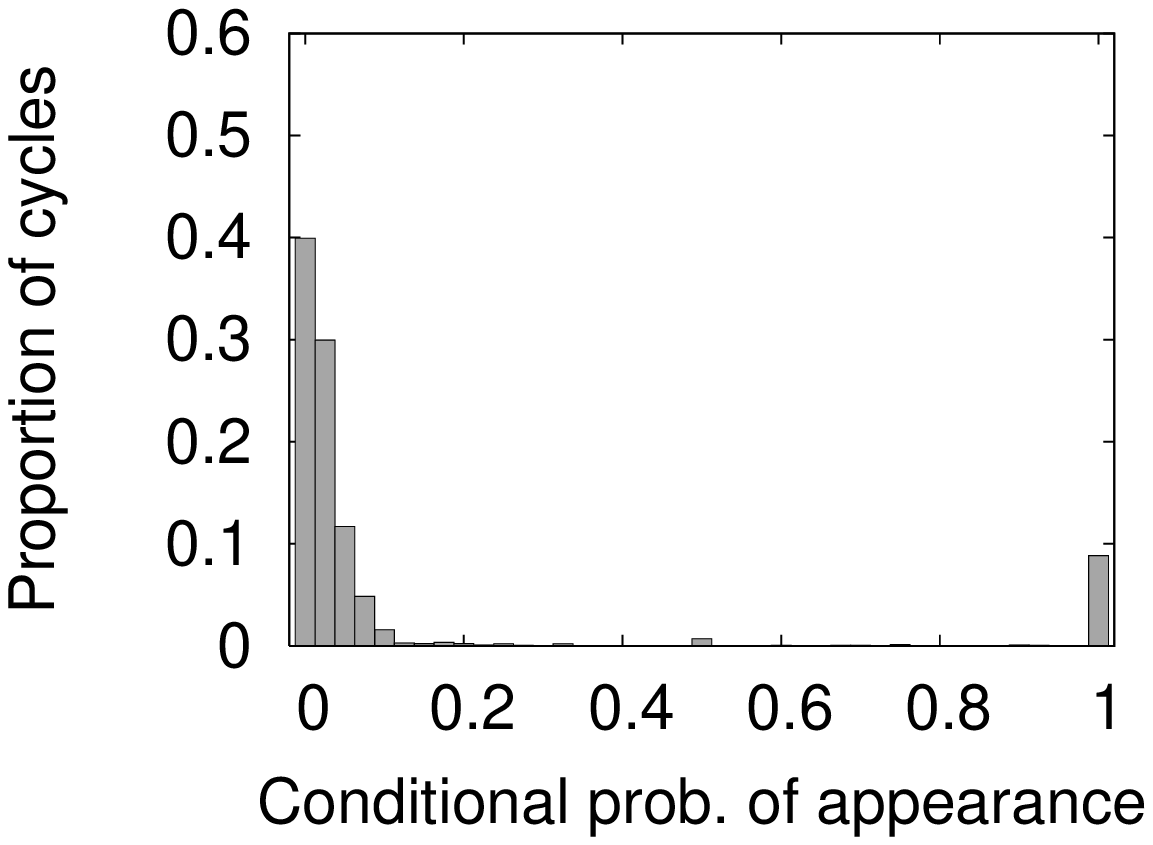}
\includegraphics[scale=0.4]{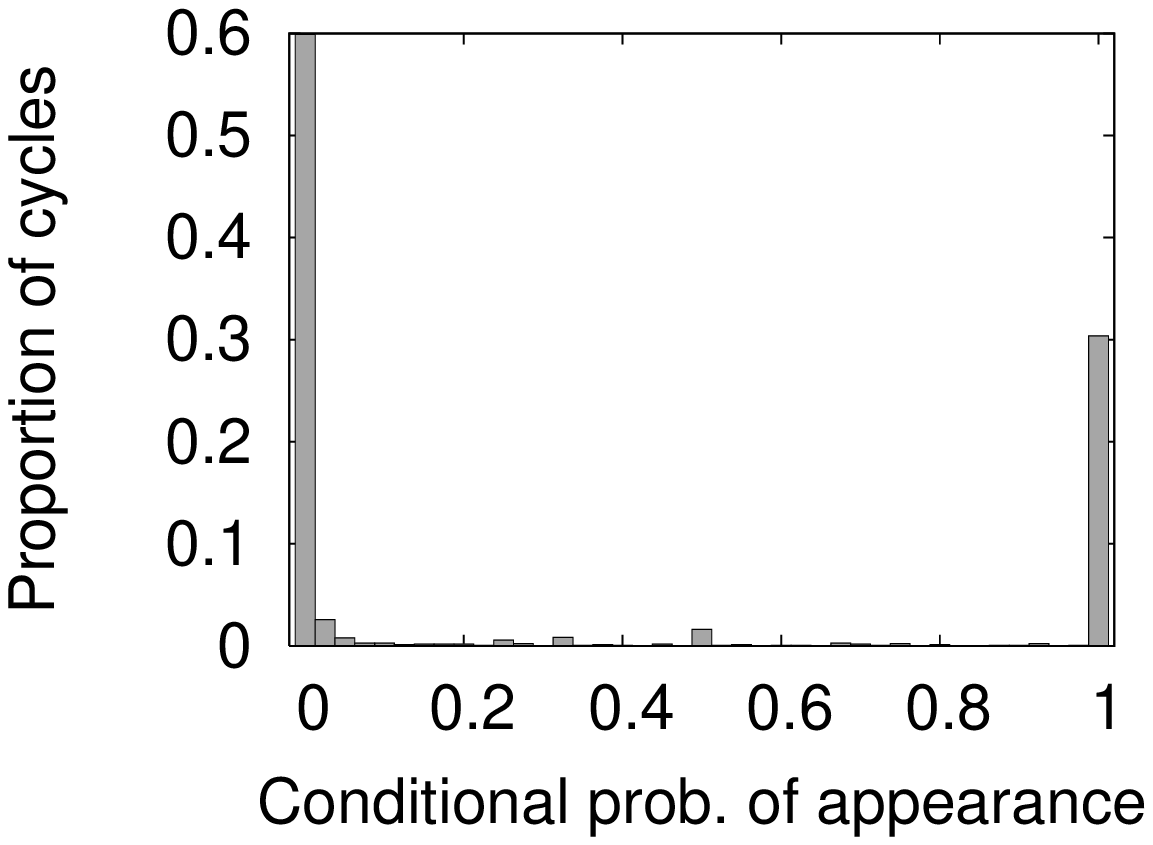}
\end{center}
\caption{Conditional appearance frequencies of cycles observed in our data set
with classic traceroute (left) and  \onelab{} (right).}
\label{fig:cycle_difference}
\end{figure}

As for loops, we also investigate 
the effect of \onelab{} on the appearance frequency of cycles.
Fig.~\ref{fig:cycle_difference} shows
that cycles that are neither
systematic nor very rare tend to disappear almost completely.
This leads to the same conclusion: load balancing is an important cause of
the appearance of cycles, and they are significantly reduced by the use of \onelab{}.

\subsection{Diamonds}
\label{impact_di}
We now compare the diamonds observed when using \onelab{}
to those observed with classic traceroute.

\begin{figure}[ht!]
\centering
\includegraphics[scale=0.4]{size_diamonds_multi_trace.eps}
\includegraphics[scale=0.4]{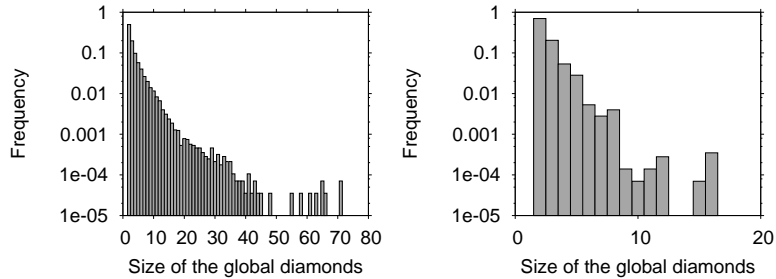}
\caption{Distribution of the size of the \globaldi{}s seen with classic traceroute (left) and \onelab{} (right).}
\label{fig_size_d_tu_multi}
\end{figure}

We observe far fewer \globaldi{}s with \onelab{}:
$52.6\%$ of the diamonds
observed with classic traceroute disappear;
conversely, new \globaldi{}s, representing $3.26\%$ of the total observed
with traceroute, appear.
Fig.~\ref{fig_size_d_tu_multi} presents
the distribution of the size of \globaldi{}s observed with \onelab{}.

We observe that, not only are there far fewer \globaldi{}s when using
\onelab{},
but their sizes are also greatly reduced.
Moreover, they are less intertwined:
$11.3\%$ of the IP addresses belong to at least one \globaldi{}'s head,
 $24.8\%$ to a core,
and $21\%$ to a tail.
The sum of these proportions is 57.1,
with overall $39.5\%$ of the addresses belonging to a \globaldi{}
(compared to $16.4\%$ of addresses in a \globaldi{}'s head,
$30.7\%$ to a core, $27.1\%$ to a tail, the sum of these proportions
being $74.2$, with $44.6\%$ of all addresses belonging to a \globaldi{},
for classic traceroute).

\begin{figure}[ht!]
\centering
\includegraphics[scale=0.4]{size_diamonds_1dst_trace.eps}
\includegraphics[scale=0.4]{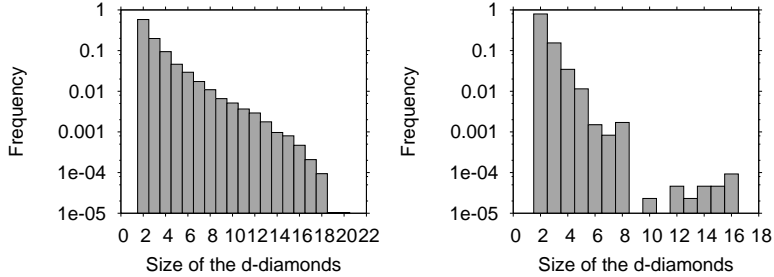}
\caption{Distribution of the size of the \ddi{}s seen with classic traceroute (left) and \onelab{} (right).}
\label{fig_size_d_tu_single}
\end{figure}

The number, size and complexity of \ddi{}s and \onedstdi{}s are also greatly reduced:
$57.1\%$ of the \ddi{}s and $56.9\%$ of the \onedstdi{}s disappear;
conversely, $2.18\%$ and $3.26\%$ of the total of \ddi{}s and \onedstdi{}s observed with
traceroute appear.
In addition, among all destinations $d$, $89.6\%$ lead to the observation of \ddi{}s
(compared to $90.2\%$ for classic traceroute),
and each such destination leads on average to the observation of $9.7$ \ddi{}s
(compared to $21.3$ for classic traceroute).
Fig.~\ref{fig_size_d_tu_single} presents
the distribution of the size of \ddi{}s observed with \onelab{}.

This comparison gives an indication of  the impact of
per-flow load balancing on our observations.
Diamonds that disappear when switching from classic traceroute to \onelab{}
are most probably induced by false links due to per-flow load balancing.
$52.6\%$ of the \globaldi{}s
observed with traceroute
disappear with \onelab{} in our data set.
This shows that diamonds are quite often (approximately half of the time) 
measurement artifacts induced by per-flow load balancing.

\medskip
Studying diamonds also makes it possible to observe per-destination
load balancing,
by studying the
differences between \globaldi{}s and \onedstdi{}s.
To understand this, we must first notice that all diamonds 
indicate {\em a priori} the existence of alternative paths between our source and some
router beyond the diamond.
This can be seen as follows:
if a given diamond is not a measurement artifact and does reflect the topology of the
network, then it represents alternative paths between its head and
its tail.
False diamonds, caused by load balancing,
also indicate the existence of alternative
paths, but not necessarily between their head and tail.
Consider the example of Fig.~\ref{fig_ex_diamonds}.
In this case, the existence of three alternative paths between $L$ and $G$
induces the appearance of diamonds.
Though there is only one path from $L$ to $D$, the diamond $(L_0, D_0)$
indicates the existence of alternative paths between some point, at its head
or before it (in this case $L$), and some other point, at its tail or after it
(in this case $G$).

Therefore, \globaldi{}s that are not \onedstdi{}s,
and similarly \globaldi{}s that are larger than corresponding \onedstdi{}s,
indicate the existence of alternative paths that can be explored only by
probing towards different destinations;
this corresponds to per-destination load balancing.
In our case, the differences between \globaldi{}s and \onedstdi{}s are similar
for traceroute and \onelab{}:
in both cases, \globaldi{}s are larger than \onedstdi{}s (this size difference is
much more pronounced for traceroute than for \onelab{}),
and \globaldi{}s that are not \onedstdi{}s appear:
there are 3,905 such \globaldi{}s for traceroute, and 3,038 for \onelab{}.
We can therefore observe a significant quantity of per-destination load balancing
in our data set.

Finally, notice that per-destination load balancing does not in itself induce
false links, and therefore does not induce false diamonds:
these are caused by the fact that probes for a same \opphroute{}
follow different paths.
A load balancer that directs packets depending on their destination,
however, will direct all probes of a given \opphroute{} on the same path,
because they all have the same destination.

\medskip
Our observations showed different things.
First, comparing diamonds obtained with the classic traceroute and
\onelab{}
shows that many diamonds seen with traceroute are measurement artifacts
caused by per-flow load balancing, and disappear when using \onelab{}.
This again shows that per-flow load balancing 
may have a strong impact on one's observations with traceroute.
Second, comparing \globaldi{}s with \onedstdi{}s allowed us to
evaluate the amount of per-destination load balancing in our
observations. 
Although this latter does not cause measurement artifacts in \opphroute{}s,
being able to detect it leads to interesting observations.

Our studies open the way to more detailed analysis of both types
of load balancing.
We note however that, for this type of studies, other measurement
strategies will probably need to be employed.
Indeed, as we already mentioned, we observed in our data set that a small number of diamonds
appear with \onelab{} but not with classic traceroute.
As explained in Sec.~\ref{loops}, this comes from the fact that some
diamonds are observed very seldom, and might therefore be observed
only with one tool and not the other.
Some of these diamonds therefore could potentially be observed with \onelab{}.
A detailed study of load balancing should address this
question.

Concerning diamonds seen with \onelab{},
we observe that \onelab{} is still subject to per-packet load balancing.
These diamonds may therefore either reflect the real topology,
or be false diamonds caused by per-packet load balancing or routing changes 
during the traceroute.
Designing tools and measurement strategies allowing the statistical
discovery of measurement artifacts caused by per-packet load balancing is
an interesting and challenging direction of work;
see Sec.~\ref{conclusion}.

\def\persistentloopsvolonelab{63.4}
\def\loopsttl{17.1}
\def\loopsttlvol{51.1}
\def\persistentloopsttl{68.6}
\def\loopsunreachable{62.0}
\def\loopsunreachablevol{19.7}
\def\cyclesunreachable{27.0}
\def\cyclesunreachablevol{3.5}
\def\cyclesforwarding{60.2}
\def\cyclesforwardingvol{91.6}
\def\loopsfake{6.8}
\def\loopsfakevol{18.6}
\def\cyclesfake{1.6}
\def\cyclesfakevol{3.0}
\def\loopsnoexp{14.0}
\def\loopsnoexpvol{10.6}
\def\cyclesnoexp{11.1}
\def\cyclesnoexpvol{1.9}

\newcommand{\longestperiodiccycle}{a week}

\section{Other causes of artifacts}
\label{other_causes}

In the previous section we have seen that most loops, most cycles, and
approximately half of the diamonds observed in traceroute measurements 
disappear when using \onelab{}.
We will now study the ones that persist.
We will see that \onelab{} provides information that
allows us to understand the causes of some of the remaining structures,
and in some cases shows that they also are measurement artifacts.

The statistics we provide in this section are in regards to the set
of structures yet unexplained by load balancing: unless stated otherwise,
the percentages and ratios are to be taken as percentages and ratios 
{\em among these remaining structures}.

\subsection{Zero-TTL forwarding}
\begin{figure}[!ht]
\begin{center}
\includegraphics[scale=0.4]{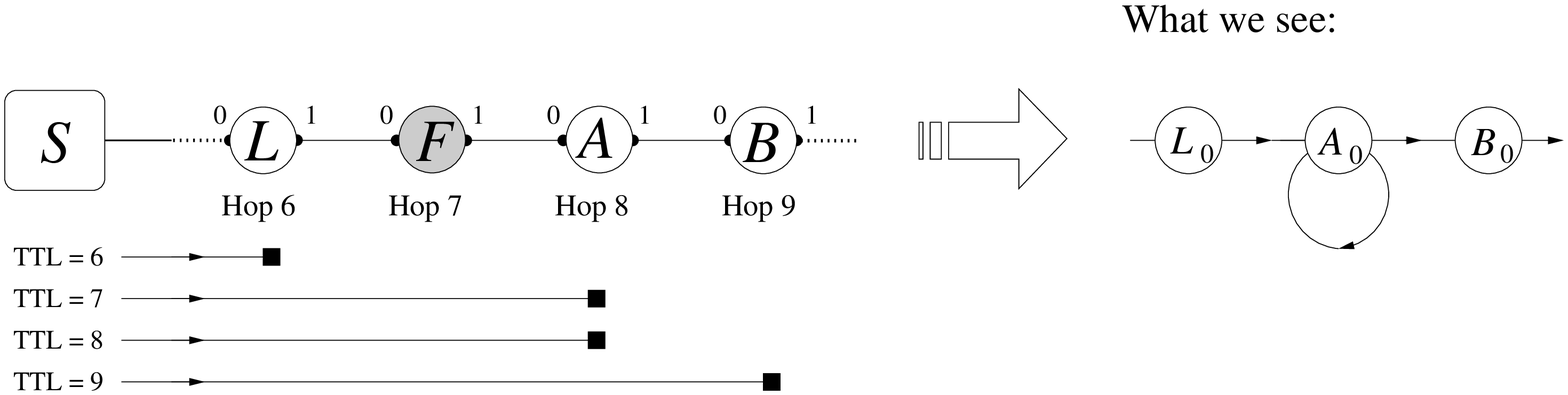}
\end{center}
\caption{Loop caused by a misconfigured router (F) that forwards packets having TTL $0$.}
\label{fig:loop-bsd}
\vspace{-3mm}\end{figure}

One explanation for loops comes from 
the \traceroute{} manual, that mentions a bug
in the standard router software of some BSD versions:
these routers decrement the TTL of packets when it is equal to one,
and forward the packet with TTL \emph{zero} (whereas a normal router would drop
the packet and generate an ICMP {\it Time Exceeded} message).
Fig.~\ref{fig:loop-bsd} shows the
consequence of the presence of such a router on a route trace.

We may validate the hypothesis of a misconfigured router with a simple experiment.
Recall that \onelab{} gives us the probe TTL: for any loop
caused by zero-TTL forwarding, the probe TTL of the first
response from the router involved in the loop would be zero. 
Fig.~\ref{fig:tracerouteoutput-loopbsd} shows how
\onelab{} would flag the loop in Fig.~\ref{fig:loop-bsd}
with a ``!T0'', indicating that the probe TTL returned for
hop 7 was zero.

\begin{figure}[!ht]
\begin{verbatim}
6  10.10.146.134  452.135 ms 
7  10.10.127.197  452.246 ms !T0 
8  10.10.127.197  450.898 ms 
9  10.10.127.37   451.499 ms 
\end{verbatim}
\caption{\onelab{} output for the example in Fig.~\ref{fig:loop-bsd}.}
\label{fig:tracerouteoutput-loopbsd}
\end{figure}

We ran this validation test on  all loops detected in our
experiments. 
Of the
persistent loop signatures we observed, $\persistentloopsttl \%$  were caused by zero-TTL forwarding.
These represent
$\loopsttl \%$ of all the loop signatures that remained with \onelab{}.
Since persistent loops represent a significant part
($\persistentloopsvolonelab \%$) of the loop instances
observed with \onelab{}, the number of loop instances that remain yet
unexplained was reduced by $\loopsttlvol \%$.

\subsection{Routing Cycles}
For cycles, the first explanation that comes to mind is that the packets
{\em do} follow a cyclic route (often called a {\em forwarding loop},
but for clarity we will avoid this term here),
and thus that the cycle is not an artifact of the measurement.
The often long span of periodic cycles, their periodic behavior, and the fact
that the observation of such cycles is typically associated
with a \opphroute{} that does not reach the destination, 
argue for this.

We hypothesize that \ip{} packets that \emph{do} follow a cyclic route
most likely do so because of a 
transient instability during routing convergence.\,\footnote{After
a topology change, routers may have inconsistent views of 
topology during some time, which is called routing convergence.}
One observation that tends to validate this hypothesis is that almost all 
periodic cycles are transient, i.e., 
packets follow a {\em normal} route for some period before and after 
we observe the cycle. Note that
this transient period
may last longer than for just one route measurement: some periodic cycles
were observed during a time span of as long as \longestperiodiccycle{}.

Moreover, we used Spring et al.'s~\cite{ally2004} technique
to verify that the identical \ip{} addresses 
which form periodic cycles really come from the same router.
This method is based on the fact that a large proportion
of routers ($52\%$ in our data set) use an internal counter to assign
ID fields to the packets they create, and increment it by $1$ every time
a packet is sent
(among the $48\%$ of remaining routers in our data set, more than
$99\%$ of them use a constant ID value set to $0$).
Packets sent within a short time period
by a router using an internal counter
have ID values that are close in regards to the $2^{16}=\text{65,536}$ possible
values of this field. 
Since routers generate far fewer
packets than they forward, this method can be applied on time windows
as large as the duration of a full \opphroute{} (typically less than 30 s).
While the observation of two packets having close ID fields
is not a proof, the accumulation of such observations
over time becomes strong evidence.
We applied this method -- when it was applicable -- to our 
periodic cycles, since \onelab{} provides the ID field of the
response packet.
We found a $100\%$ positive 
response, which shows that the periodic cycles we
observe do in fact correspond to the actual routing.
Routing cycles therefore represent $\cyclesforwarding \%$ of the cycle
signatures, and $\cyclesforwardingvol \%$ of the cycle instances.

Finally, $43\%$ of the \opphroute{}s with
periodic cycles actually reached the destination. According to 
a previous study of packet-level measurements in a large ISP 
backbone~\cite{hengartner:02}, most routing cycles last for less
than $10$ seconds. This time is less than the average time of $30$ seconds
needed to measure a route, so it seems reasonable
to assume that packets may be caught in a cycle for some time, and then reach the
destination. However, the study also found examples of routing cycles
that persisted for $10$--$60$ seconds, which may be the explanation
for cycles that do not end during our measurements. 

\subsection{Interrupted routes}
Another explanation for loops, which also applies to cycles, comes from
ICMP `unreachable' responses. Some routers, when realizing that a packet cannot
be delivered to its destination for some reason, may issue a special ICMP response,
such as an ICMP {\em Host Unreachable}, {\em Network Unreachable},
or {\em Source Quench} (among others), and discard the probe.
When receiving such responses,
both classic \traceroute{} and \onelab{} consider that the route measurement
has ended.
Now, these routers  may send these special responses even after
having forwarded one or several probes. In this case, the router will appear
twice on the \opphroute{}: once when it sends the normal ICMP
{\em Time Exceeded} message, and once later when it sends the ICMP `unreachable' packet.

We observe empirically that, most of the time in these cases, the router sends the unreachability
message just after the {\em Time Exceeded}, leading to the
observation of a loop. But sometimes it may also forward several probe
packets between these two events, thus creating a cycle.

Tracking these events is easy, as \onelab{} displays the type of ICMP received
as answers, which allows us to filter out these messages.
Interrupted routes represent $\loopsunreachable \%$ of the loop signatures,
and $\cyclesunreachable \%$ of the cycle signatures.
In terms of instances, these numbers become respectively
$\loopsunreachablevol \%$ and $\cyclesunreachablevol \%$.
The portion of instances is comparatively low since this type of measurement
artifact is by essence occasional: {\em most} routers do not issue this 
type of message very often, and only a couple of loops and cycles that 
were caused by interrupted routes were observed persistently.

\subsection{Fake IPs in return packets}
Our last hypothesis for the presence of loops comes from the observation of the
persistent \nloops.
Some subnetworks are known to be impervious to traceroute measurements
because they are behind
NAT boxes and firewalls. 
In these networks the routers 
\emph{replace} the Source Address field of 
the {\it Time Exceeded} packets on the way back to the probing machine,
making them appear to come from their border gateway.

To validate this hypothesis, we tried to get some evidence that the 
consecutive replies from seemingly identical \ip{} addresses actually came
from different routers. The simplest way is to compare the TTL of the
ICMP packets they send back: if the TTLs are different, then the responses
likely came from different routers.
This is even more obvious if these TTLs
are always decreasing with distance
(which indicates that the answering routers are in fact aligned on a route.)

We call \emph{fake loops} the loops that are caused by this kind
of Source Address  replacement.
We observed that $\loopsfake \%$ of the loops signatures and
$\loopsfakevol \%$ of the loop instances came from such substituted \ip{}s.
The fake loops we detected were mostly persistent or systematic.
One interesting result is that most of the persistent loops
detected in our experiment that remained unexplained (i.e., not caused by
zero-TTL forwarding) were determined to be fake loops.
Another interesting result is that all persistent $n$-loops in our data set were fake.

The case of cycles caused by fake \ip{}s is more puzzling, since the explanation
based on protected subnetworks seems less plausible when observing
non-consecutive responses with the same \ip{} address.
There are far fewer: only
$\cyclesfake \%$ of the cycle signatures and $\cyclesfakevol \%$ of their
instances seemed to be caused by it. 

\subsection{Summary}

By considering per-flow load balancing,
zero-TTL forwarding, 
routing cycles, interrupted routes, or
fake IPs in return packet as causes for artifacts,
we were able to explain
roughly half of the diamonds,
and more than
$95 \%$ of loops and cycles in our data set.
 
Table~\ref{tab:summary} presents the details of the proportions
of loops (signatures and instances), cycles (signatures and instances)
and diamonds (\globaldi{}s and \onedstdi{}s)
caused in our data set by each phenomenon.
In contrast to the other numbers presented in this section,
the percentages in this table are with respect to {\em all} the
structures observed in our data set:
we do not restrict ourselves to the structures that remain with \onelab{}.
Since not all phenomena are possible causes of appearance for all structures,
a dash `-' indicates that a
phenomenon does not apply to a structure.

Though our data are not representative,
this illustrates how \onelab{} helps understand and detect traceroute
measurement artifacts.
Regarding the unexplained structures,
we suspect, after some preliminary studies,
that most are artifacts caused
by per-packet load balancing.

\FPeval\loopLB{round(disappearloop-appearloop,2)}
\FPeval\loopLBvol{round(disappearloopvol-appearloopvol,2)}
\FPeval\cycleLB{round(disappearcycle-appearcycle,2)}
\FPeval\cycleLBvol{round(disappearcyclevol-appearcyclevol,2)}

\FPeval\loopttl{round(loopsttl*(100-loopLB)/100,2)}
\FPeval\loopttlvol{round(loopsttlvol*(100-loopLBvol)/100,2)}
\FPeval\loopunreach{round(loopsunreachable*(100-loopLB)/100,2)}
\FPeval\loopunreachvol{round(loopsunreachablevol*(100-loopLBvol)/100,2)}
\FPeval\cycleunreach{round(cyclesunreachable*(100-cycleLB)/100,2)}
\FPeval\cycleunreachvol{round(cyclesunreachablevol*(100-cycleLBvol)/100,2)}
\FPeval\cycleforward{round(cyclesforwarding*(100-cycleLB)/100,2)}
\FPeval\cycleforwardvol{round(cyclesforwardingvol*(100-cycleLBvol)/100,2)}
\FPeval\loopfake{round(loopsfake*(100-loopLB)/100,2)}
\FPeval\loopfakevol{round(loopsfakevol*(100-loopLBvol)/100,2)}
\FPeval\cyclefake{round(cyclesfake*(100-cycleLB)/100,2)}
\FPeval\cyclefakevol{round(cyclesfakevol*(100-cycleLBvol)/100,2)}
\FPeval\loopnoexp{round(loopsnoexp*(100-loopLB)/100,2)}
\FPeval\loopnoexpvol{round(loopsnoexpvol*(100-loopLBvol)/100,2)}
\FPeval\cyclenoexp{round(cyclesnoexp*(100-cycleLB)/100,2)}
\FPeval\cyclenoexpvol{round(cyclesnoexpvol*(100-cycleLBvol)/100,2)}

\def\tabw{0.04\textwidth}

\begin{table}
\begin{center}
\small
\scalebox{0.6}{
\begin{tabular}[c]{| p{0.25\textwidth} | r@{.}l | r@{.}l | r@{.}l | r@{.}l | r@{.}l | r@{.}l |}
 \cline{2-13}
   \multicolumn{1}{c|}{}
 & \multicolumn{4}{|c|}{Diamonds}
 & \multicolumn{4}{|c|}{Loops}
 & \multicolumn{4}{|c|}{Cycles}
\\ \cline{2-13}
   \multicolumn{1}{c|}{}
 & \multicolumn{2}{|c|}{global} 
 & \multicolumn{2}{|c|}{$1$-dst}
 & \multicolumn{2}{|c|}{sign.} 
 & \multicolumn{2}{|c|}{inst.}
 & \multicolumn{2}{|c|}{sign.} 
 & \multicolumn{2}{|c|}{inst.}
\\ \hline
Per-flow load bal.
 & 52&6\ 
 & 56&9\ 
 & 87&1\ 
 & 79&7\ 
 & 68&5\ 
 & 38&4\ 
\\ \hline
Zero-TTL fwd.
 & \multicolumn{2}{|c|}{-}
 & \multicolumn{2}{|c|}{-}
 &  2&21
 & 10&4\ 
 & \multicolumn{2}{|c|}{-}                         
 & \multicolumn{2}{|c|}{-}
\\ \hline
Routing cycles
 & \multicolumn{2}{|c|}{-}
 & \multicolumn{2}{|c|}{-}
 & \multicolumn{2}{|c|}{-}                         
 & \multicolumn{2}{|c|}{-}
 & 19&0\ 
 & 56&4\ 
\\ \hline
Interrup. routes
 & \multicolumn{2}{|c|}{-}                         
 & \multicolumn{2}{|c|}{-}
 & 8&00
 & 4&00
 & 8&51
 & 2&16
\\ \hline
Fake IP addresses
 & \multicolumn{2}{|c|}{-}                         
 & \multicolumn{2}{|c|}{-}
 & 0&88
 & 3&78
 & 0&50
 & 1&85
\\ \hline
Unknown 
 & 47&4\ 
 & 43&1\ 
 &  1&81
 &  2&15
 &  3&50
 &  1&17
\\ \hline
Total
 & 100&00
 & 100&00
 & 100&00             
 & 100&00
 & 100&00
 & 100&00
\\ \hline
\end{tabular}
}
\end{center}
\caption{Summary of identified measurement artifacts,
showing the proportion of observed structures in our data that are
attributed to each cause.  All values are percentages.}
\label{tab:summary}
\end{table}

\section{Related work}
\label{Related Work}\label{related_work}

We have earlier published a short version of this paper~\cite{augustin2006paris}.
The present paper extends this preliminary work in several ways:
the data set used here is much larger; we present here much more detailed
analysis of loops and cycles; and the discussion on diamonds (Sec.~\ref{diamonds}
and~\ref{impact_di}) is almost completely new.

The principal variants on Jacobson's
traceroute~\cite{jacobson1989traceroute} are Gavron's NANOG
traceroute~\cite{gavron1995nanog},
Eddy's prtraceroute~\cite{eddy1994prtraceroute},
and Toren's
tcptraceroute~\cite{toren2001tcptraceroute}.  NANOG traceroute and prtraceroute both label
IP addresses with the numbers of the ASes to which they belong.  Tcp\-trace\-route sends
TCP probes (rather than the classic UDP or ICMP Echo probes) using
Destination Port 80 to emulate web traffic and thus more easily
traverse firewalls.  As described in Sec.~\ref{a new traceroute}, this
has the effect of maintaining a constant flow identifier.  No
prior work however has looked at the use of this feature to
avoid traceroute measurement artifacts.

Although there is an extensive literature on internet maps, and much
work that uses \traceroute{}, there have been few studies of artifacts
as seen from the perspective of traceroute. 
Moors~\cite{moors2004streamlining} suggests that encoding traceroute
probe packet identifiers in the Source Port field would
allow the use of destination ports associated with classical services
(e.g., HTTP or SMTP).
This would cause all network elements, including load balancers,
 to handle these packets identically to the packets of normal
network traffic.  In doing so, he correctly identifies the problem that
traceroute may not report routes that normal packets usually follow.
However, the proposed
solution still alters the five-tuple, 
and a tool built in this way would still suffer from load balancing
and hence would present the same measurement artifacts as classic traceroute.

Paxson's work on end-to-end 
routing behavior in the internet~\cite{paxson1999endtoend} uses 
\traceroute{} to study routing dynamics, including ``routing pathologies''. 
Although some of these path\-ol\-ogies do relate to the
structures
we discuss in this paper (for instance, ``routing loops'' are one cause of 
``cycles'', and ``fluttering'' is one cause of ``diamonds''), his work focuses on the 
routing aspect of his observations and not on \traceroute{}'s deficiencies.
Teixeira et al.~\cite{teixeira2003search} examine inaccuracies
introduced into ISP maps obtained by Rocketfuel~\cite{spring2002measuring}.  
Their paper quantifies differences between the true
and the measured topologies, and identifies routing changes in the
midst of individual traceroutes as being responsible for a 
portion of the false links in the measured topologies,
but does not touch on load balancing.

Some topology inference systems based on traceroute
acknowledge the problem that traceroute may report several interfaces
at a same hop on a given path and thus may infer false links.
They handle these problems
in different ways.  Huffaker et al.\ recognize the problem for
\textit{skitter}~\cite{huffaker2002topology}, but they do not report a
solution.  In practice, 
skitter sends three probes per hop and the routes reported by
the \textit{arts++} tool for reading skitter
data consist of the 
first address obtained for each hop.
With \textit{Rocketfuel}~\cite{spring2002measuring}, Spring et al.\
attribute a lower confidence level to links inferred from hops that
respond with multiple addresses. Still, they include all these links
in the database from which they construct a network's
topology.  Their hope is that a subsequent alias resolution step will eliminate
at least some of the false links. However, this only works if load
balancing takes place over multiple links between the same pair of
routers.

\section{Conclusion}\label{conclusion}

In this paper, we identified and characterized three structures 
appearing frequently in traceroute measurements: loops, cycles,
and diamonds.
We explained how these structures, some of which are often
attributed to routing dynamics or pathologies,
may rather be measurement artifacts, induced notably by load balancing.
We designed the \onelab{} tool, a new traceroute that finds accurate
routes under per-flow load balancing and proposed rigorous methods 
for checking the cause of each structure.
By conducting side-by-side experiments with classic traceroute and \onelab{},
we were able to show that most of these structures appearing in our
traceroute traces are in fact measurement artifacts, and are avoided with \onelab{}.
Though our measurements are made from one source only and cannot be
considered as statistically representative of what one can expect in the internet in
general, this shows that per-flow load balancing can have a strong
impact on one's observations,
and that the view obtained with \onelab{} is more accurate.

\medskip
This work may inspire future research in many directions. 
First, our experiments exposed other, rarer, structures
than those described here, that also deserve attention: we observed 
cliques (sets of interfaces all linked to each other),
dense regions, and IP addresses with a very high number of links.
These structures are surprising, and may also be measurement
artifacts.

Second, we believe that  \onelab{} can be further improved.
We are working on an algorithm to automatically
discover all paths between a source and a destination in the
presence of per-flow load balancing.
We are also developing techniques to uncover
accurate routes in the presence of
per-packet load balancing.

Finally, an important and interesting
extension of this work is to perform measurements,
both with classic traceroute and \onelab{}, from several sources.
This would make it possible to quantify the amount of traceroute
measurement artifacts in the internet in general.
We believe that new artifacts and features of interest will be observable when 
measurements originate from multiple sources.
Moreover, one of the main conclusions of this work being that \onelab{}
provides a much truer view of the topology than
existing tools,
conducting measurements with \onelab{} will
yield data of higher quality for  studying the topology
of the internet.
 Studying the impact of using \onelab{} on the observed
 topological properties is a very promising direction.

\section*{Acknowledgments}

The \onelab{} tool was developed with financial support from
the French CNRS, as part of its contribution to the European
Commission-sponsored \textit{OneLab} project.  The research using the
tool was conducted with financial support from the French MNRT ACI
\textit{S\'ecurit\'e et Informatique} 2004 grant, through the
\textit{METROSEC} project,
and from the French ANR \textit{Jeunes chercheuses et jeunes chercheurs}
2005 grant, through the \textit{AGRI} project.

We thank Ehud Gavron and Dave Andersen for their thoughtful comments.
We also thank
Neil Spring and Ratul Mahajan for their explanations of  how
Rocketfuel handles hops that respond with multiple interfaces. 


\bibliographystyle{elsart-num}
\bibliography{bibliography}

\begin{thebibliography}{10}
\expandafter\ifx\csname url\endcsname\relax
  \def\url#1{\texttt{#1}}\fi
\expandafter\ifx\csname urlprefix\endcsname\relax\def\urlprefix{URL }\fi

\bibitem{jacobson1989traceroute}
V.~Jacobson, traceroute, the most recent version is available at:
  \url{ftp://ftp.ee.lbl.gov/traceroute.tar.gz} (February 1989).

\bibitem{govindan2000heuristics}
R.~Govindan, H.~Tangmunarunkit, Heuristics for internet map discovery, in:
  Proc. IEEE Infocom, 2000.

\bibitem{huffaker2002topology}
B.~Huffaker, D.~Plummer, D.~Moore, k~claffy, Topology discovery by active
  probing, in: Proc. Symposium on Applications and the Internet, 2002.

\bibitem{spring2002measuring}
N.~Spring, R.~Mahajan, D.~Wetherall, Measuring {ISP} topologies with
  {R}ocketfuel, in: Proc. ACM SIGCOMM, 2002.

\bibitem{Cheswick00}
B.~Cheswick, H.~Burch, {Internet Mapping Project},
  {{http://cm.bell-labs.com/who/ches/map/index.html}} (2000).

\bibitem{dimes}
Y.~Shavitt, E.~Shir, {DIMES}: Let the internet measure itself, ACM SIGCOMM
  Computer Communication Review 35~(5) (2005) 71 -- 74.

\bibitem{faloutsos99powerlaw}
M.~Faloutsos, P.~Faloutsos, C.~Faloutsos, On power-law relationships of the
  internet topology, in: Proc. ACM SIGCOMM, 1999.

\bibitem{magoni2005mapping}
D.~Magoni, M.~Hoerdt, Internet core topology mapping and analysis, ACM SIGCOMM
  Computer Communications 28~(5) (2005) 494--506.

\bibitem{moors2004streamlining}
T.~Moors, Streamlining traceroute by estimating path lengths, in: Proc. IEEE
  Workshop on IP Operations and Management, 2004.

\bibitem{augustin2006paris}
B.~Augustin, X.~Cuvellier, B.~Orgogozo, F.~Viger, T.~Friedman, M.~Latapy,
  C.~Magnien, R.~Teixeira, Avoiding traceroute anomalies with paris
  traceroute., in: Proc. ACM SIGCOMM Internet Measurement Conference, 2006.

\bibitem{gavron1995nanog}
E.~Gavron, {NANOG} traceroute, the most recent version is available at:
  \url{ftp://ftp.login.com/pub/software/traceroute/} (May 1995).

\bibitem{stevens1994traceroute}
W.~R. Stevens, {TCP}/{IP} Illustrated, Volume 1: The Protocols, Addison-Wesley,
  1994, Ch. 8, Traceroute Program.

\bibitem{toren2001tcptraceroute}
M.~Toren, tcptraceroute, see \url{http://michael.toren.net/code/tcptraceroute/}
  (April 2001).

\bibitem{ip-rfc791}
J.~Postel, Internet protocol, IETF RFC 791 (September 1981).

\bibitem{routers-rfc1812}
F.~Baker, Requirements for {IP} {V}ersion 4 routers, IETF RFC 1812 (June 1995).

\bibitem{mao2003astr}
Z.~M. Mao, J.~Rexford, J.~Wang, R.~Katz, Towards an accurate as-level
  traceroute tool, in: Proc. ACM SIGCOMM, 2003.

\bibitem{rfc792}
J.~Postel, Internet control message protocol, IETF RFC 791 (September 1981).

\bibitem{ospf-rfc2328}
J.~Moy, {OSPF Version~2}, IETF RFC 2328 (April 1998).

\bibitem{isis-rfc1195}
R.~Callon, {Use of OSI IS--IS for Routing in TCP/IP and Dual Environments},
  IETF RFC 1195 (December 1990).

\bibitem{quoitin03interdomain}
B.~Quoitin, S.~Uhlig, C.~Pelsser, L.~Swinnen, O.~Bonaventure, Interdomain
  traffic engineering with {BGP}, IEEE Communication Magazine 41~(5) (2003)
  122--128.

\bibitem{cisco-lb}
Cisco, How does load balancing work?, see
  \url{http://www.cisco.com/en/US/tech/tk365/technologies_tech_note09186a00800%
94820.shtml}.

\bibitem{juniper-lb}
Juniper, Configuring load-balance per-packet action, from the JUNOS 7.0 Policy
  Framework Configuration Guideline, see
  \url{http://www.juniper.net/techpubs/software/junos/junos70/swconfig70-polic%
y/html/policy-actions-config11.html}.

\bibitem{cisco-pf}
Cisco, Cisco 7600 {S}eries {R}outers {C}ommand {R}eferences, from the Cisco
  Documentation.

\bibitem{govindan2002estimating}
R.~Govindan, V.~Paxson, Estimating router {ICMP} generation delays, in: Proc.
  of Passive and Active Measurement Workshop, 2002.

\bibitem{bellovin2002nat}
S.~Bellovin, A technique for counting {NATted} hosts, in: Proc. ACM SIGCOMM
  Internet Measurement Workshop, 2002.

\bibitem{xia2005}
J.~Xia, L.~Gao, T.~Fei, Flooding attacks by exploiting persistent forwarding
  loops, in: Proc. ACM SIGCOMM Internet Measurement Conference, 2005.

\bibitem{mao:04}
Z.~M. Mao, D.~Johnson, J.~Rexford, J.~Wang, R.~H. Katz, Scalable and accurate
  identification of {AS}-level forwarding paths, in: Proc. IEEE Infocom, 2004.

\bibitem{special-rfc3330}
{IANA}, Special-use {IPv4} addresses, IETF RFC 3330 (September 2002).

\bibitem{ally2004}
N.~Spring, M.~Dontcheva, M.~Rodrig, D.~Wetherall, How to resolve {IP} aliases,
  UW CSE Technical Report~(04-05-04).

\bibitem{hengartner:02}
U.~Hengartner, S.~B. Moon, R.~Mortier, C.~Diot, Detection and analysis of
  routing loops in packet traces, in: Proc. ACM SIGCOMM Internet Measurement
  Workshop, 2002.

\bibitem{eddy1994prtraceroute}
R.~Eddy, prtraceroute, the most recent version is part of the Internet Systems
  Consortium's IRRToolSet, available from: \url{http://www.isc.org/} (August
  1994).

\bibitem{paxson1999endtoend}
V.~Paxson, End-to-end internet packet dynamics, IEEE/ACM Trans. Networking
  7~(3) (1999) 277--292.

\bibitem{teixeira2003search}
R.~Teixeira, K.~Marzullo, S.~Savage, G.~M. Voelker, In search of path diversity
  in {ISP} networks, in: Proc. ACM SIGCOMM Internet Measurement Conference,
  2003.

\end{thebibliography}

\end{document}